\documentclass[12pt]{article}
\usepackage{amsmath}
\usepackage{amsfonts}
\usepackage{amssymb}
\usepackage{float}
\usepackage{graphicx}
\restylefloat{table}
\usepackage{color}
\usepackage{bbm}
\usepackage{appendix}

\newlength{\abstractwidth}

\usepackage{epsf}
\usepackage{cancel}

\renewcommand{\thanks}[1]{\footnote{#1}}

\renewcommand{\theequation}{\thesection.\arabic{equation}}
\newcommand{\bea}{\begin{eqnarray}}
\newcommand{\eea}{\end{eqnarray}}
\newcommand{\ee}{\end{equation}}
\newcommand{\be}{\begin{equation}}

\newcommand{\ea}{\end{array}}
\newcommand{\bac}{\begin{array}{c}}
\newcommand{\bacc}{\begin{array}{cc}}
\newcommand{\barcl}{\begin{array}{r@{}c@{}l}}
\newcommand{\brcl}{\begin{array}{rcl}}
\newcommand{\bdm}{\begin{displaymath}}
\newcommand{\edm}{\end{displaymath}}
\newcommand{\nn}{\nonumber}

\def\cN{{\cal N}}

\def\Re{{\rm Re}}
\def\Im{{\rm Im}}

\def\no{\nonumber}

\def\fig#1{fig.~{\ref{#1}}}

\def\spa#1.#2{\left\langle#1\,#2\right\rangle}
\def\spb#1.#2{\left[#1\,#2\right]}

\newcommand{\ta}{\ensuremath{a}}

\newcommand{\tx}{\ensuremath{{x}}}

\newcommand{\ti}{\ensuremath{{I}}}

\usepackage{color}

\renewcommand{\theequation}{\thesection.\arabic{equation}}
\numberwithin{equation}{section}

\def\no{\nonumber}

\def\Tr{{\rm Tr}}
\def\tree{{\rm tree}}

\def\rf{{\rm f}}
\def\rff{{\tilde \rf}}

\def\allphi#1{{\lowercase{ #1}}}
\def\specphi#1{{\lowercase{ #1}}}

\def\SYMcoupling{{\rm g}}

\begin{document}
 
\textwidth 170mm
\textheight 230mm
\topmargin -1cm
\oddsidemargin-0.8cm \evensidemargin-0.8cm
\topskip 9mm
\headsep9pt

\overfullrule=0pt
\parskip=2pt
\parindent=12pt
\headheight=0in \headsep=0in \topmargin=0in \oddsidemargin=0in

\vspace{ -3cm} \thispagestyle{empty} \vspace{-1cm}

\begin{flushright}
AEI-2014-033; CERN-PH-TH/2014-141; IGC-14/8-1
\end{flushright}

\begin{center}

{\Large \bf Scattering amplitudes in $\cN=2$ Maxwell-Einstein and Yang-Mills/Einstein supergravity}

\bigskip

\medskip

{\large Marco Chiodaroli${}^{a}$, Murat G\"{u}naydin${}^{b}$, Henrik Johansson${}^{c}$ and Radu Roiban${}^{b}$}

\bigskip

${}^a${ Max-Planck-Institut f\"ur Gravitationsphysik \\ 
             Albert-Einstein-Institut\\
Am M\"uhlenberg 1, 14476 Potsdam, Germany\\
\texttt{mchiodar$\,$@aei.mpg.de}}

\medskip

${}^b${ Institute for Gravitation and the Cosmos \\
 The Pennsylvania State University \\
University Park PA 16802, USA \\ 
\texttt{murat,radu$\,$@phys.psu.edu}}

\medskip

${}^c${ Theory Division,\\ 
Physics Department, CERN \\
CH-1211 Geneva 23, Switzerland\\
\texttt{henrik.johansson$\,$@cern.ch}
}

\end{center}

\smallskip

\begin{abstract}

\medskip

We expose a double-copy structure in the scattering amplitudes of the generic Jordan family of $\cN=2$ 
Maxwell-Einstein and Yang-Mills/Einstein supergravity theories in four and five dimensions. 
The Maxwell-Einstein supergravity amplitudes are obtained through the color/kinematics duality 
as a product of two gauge-theory factors; one originating from pure $\cN=2$ super-Yang-Mills theory 
and the other from the dimensional reduction of a bosonic higher-dimensional pure Yang-Mills theory. 
We identify a specific symplectic frame in four dimensions for which the on-shell fields and amplitudes 
from the double-copy construction can be identified with the ones obtained from the supergravity 
Lagrangian and Feynman-rule computations. 
The Yang-Mills/Einstein supergravity theories are obtained by gauging a compact subgroup 
of the isometry group of their Maxwell-Einstein counterparts. 
For the generic Jordan family this process is identified with the introduction of 
cubic scalar couplings on the bosonic gauge-theory side, which through the double copy are responsible for the
non-abelian vector interactions in the supergravity theory. As a demonstration of the power of this structure, 
we present explicit computations at tree-level and one loop. 
The double-copy construction allows us to obtain compact expressions for the supergravity superamplitudes,  
which are naturally organized as polynomials in the gauge coupling constant.    \end{abstract}

\baselineskip=16pt
\setcounter{equation}{0}
\setcounter{footnote}{0}

\newpage

\tableofcontents

\newpage

\section{Introduction}

Perturbative $S$ matrices of gravitational theories in flat spacetime 
enjoy a particularly simple structure that remains highly obscure at the level of the
Lagrangians and in covariant computations based on Feynman diagrams. 
Evidence for this structure
was first brought to light through the Kawai-Lewellen-Tye (KLT) relations~\cite{KLT, KLT2}, which demonstrate that the information encoded in a gauge-theory tree-level $S$ matrix is sufficient to construct a tree-level gravity $S$ matrix.

More recently, Bern, Carrasco and Johansson (BCJ) proposed a set of Lie-algebraic relations for the kinematic 
building blocks of gauge-theory loop-level amplitudes~\cite{BCJ, BCJLoop}, which mirror analogous relations obeyed 
by the corresponding color building blocks. Their existence mandates that 
gravity amplitudes can be obtained from gauge-theory amplitudes through a double-copy construction. 
The double copy trivializes the construction of gravity loop-level integrands when the corresponding gauge amplitudes are available in a 
presentation that manifestly satisfies such  duality between color and kinematics. 
The existence of amplitudes presentations that obey color/kinematics duality has been conjectured for broad 
classes of Yang-Mills (YM) theories to all loop orders and multiplicities~\cite{BCJLoop}. 

There is by now substantial evidence for the duality at tree level. Multiple versions of duality-satisfying tree amplitudes in pure super-Yang-Mills (sYM) theories, 
including arbitrary number of legs and in $D$ dimensions~\cite{KiermaierTalk, BjerrumBohr:2010hn, Mafra:2011kj}, have been 
constructed~\cite{Tye:2010dd, Monteiro:2011pc,Broedel:2011pd, Johansson:2013nsa,Litsey:2013jfa,Fu:2014pya,Naculich:2014rta}. 
The duality enforces the so-called BCJ relations between color-ordered tree amplitudes, which have been proven 
from both string-~\cite{BjerrumBohr:2009rd,Stieberger:2009hq} and field-theory~\cite{Feng:2010my,Chen:2011jxa, Cachazo:2012uq} perspectives. 
Duality-satisfying structures are known to naturally arise from the field-theory limit of string theory using the pure-spinor formalism \cite{Mafra:2011kj,Mafra:2009bz,Mafra:2011nw, Broedel:2013tta, Stieberger:2014hba, Mafra:2014oia}. 
Attempts toward a Lagrangian understanding of the duality are ongoing~\cite{Bern:2010yg,Tolotti:2013caa}, 
and the study of the kinematic Lie algebra underlying 
the duality has made significant advances~\cite{Monteiro:2011pc, BjerrumBohr:2012mg,Fu:2012uy,Boels:2013bi,Monteiro:2013rya}. Color/kinematics 
duality played a key role in the advent of
the scattering equations~\cite{Cachazo:2012uq, Cachazo:2012da,Cachazo:2013gna, Cachazo:2013hca, Cachazo:2013iea,Naculich:2014naa}. 
That the duality is not limited to YM theories, but applies to certain Chern-Simons-matter theories, was observed in refs.~\cite{Bargheer:2012gv,Huang:2012wr,Huang:2013kca}.

Amplitude presentations with manifest color/kinematics duality at loop-level were first constructed 
for the maximal sYM theory \cite{Bern:2010tq,Carrasco:2011mn,Bern:2012uf} and have played a critical role in enabling state-of-the-art multiloop computations in (ungauged)  $\cN \ge4$ supergravities 
\cite{Bern:2011rj,BoucherVeronneau:2011qv, Bern:2012cd,Bern:2012gh,Bern:2013qca,Bern:2013uka}. 
These calculations, as well as earlier ones, have exposed unexpected ultraviolet (UV) cancellations. 
The four-point amplitudes of $\cN=8$ supergravity were shown to be manifestly UV-finite through four loops~\cite{Bern:2007hh,Bern:2008pv,Bern:2009kd,Bern:2012uf}, and later work proved that the $\cN=8$ theory cannot have a divergence before seven loops in four dimensions~\cite{Bossard:2010dq,Bossard:2010bd,Beisert:2010jx,Bossard:2011tq}. Interestingly, 
finiteness until this loop order agrees with earlier na\"{\i}ve power-counting based on 
the assumed existence of an off-shell ${\cal N} = 8$ superspace \cite{Grisaru:1982zh}. The potential existence of  
seven-loop divergences has been suggested from several perspectives, including an analysis of
string theory dualities~\cite{GRV2010}, a first-quantized
world-line approach~\cite{FirstQuantized}, and light-cone
supergraphs~\cite{KalloshRamond}.  However, it has also
been argued that the theory may remain finite beyond seven
loops~\cite{FiniteArgue}.
 Pure $\cN=4$ supergravity has been shown to be UV-finite at three loops \cite{Bern:2012cd,Bern:2012gh},
 and to diverge at four loops \cite{Bern:2013uka}. 
 The authors of ref.~\cite{Bern:2013uka} suggested that the appearance of the four-loop divergence should be related to the quantum anomaly of a $U(1)$ subgroup of 
 the $SU(1,1)$ global symmetry of the theory \cite{Carrasco:2013ypa}.  

More recent work studied duality-satisfying presentations for one-loop gauge-theory amplitudes, with and without adjoint matter, 
for $\cN\leq2$ supersymmetric theories~\cite{Carrasco:2012ca, Bern:2013yya, Nohle:2013bfa, Ochirov:2013xba}.\footnote{See also refs.~\cite{Bjerrum-Bohr:2013iza,Mafra:2012kh,Tourkine:2012vx,Green:2013bza} for explicit one-loop calculations related to color/kinematics duality.} In ref.~\cite{Carrasco:2012ca} it was first shown that there exists double-copy structures for particular classes of gravity theories with $\cN < 4$ that can be constructed as field-theory orbifolds of $\cN=8$ supergravity. To access a broader spectrum of supergravities with reduced $\cN < 4$ supersymmetry it is necessary 
to generalize color/kinematics duality by including matter fields that do not transform in the adjoint representation; the bi-fundamental \cite{Chiodaroli:2013upa} and fundamental~\cite{Johansson:2014zca} representations are natural generalizations. 
In these cases, each line of a gauge-theory graph has a definite gauge-group representation through its color factor, and, generically, proper double copies are obtained by matching up pairs of numerator factors that belong to conjugate (or, alternatively, identical) color representations~\cite{Chiodaroli:2013upa, Johansson:2014zca}.
      

Further generalizations (deformations)  of supergravity theories are achieved by  gauging.  In supergravity theories with abelian vector fields, 
one can gauge a subgroup of the global symmetry group, and/or $R$-symmetry group, such that some of the vector fields become gauge fields of the gauged group. 
This (sub)set of vector fields must transform in the adjoint representation of the corresponding gauge group.
The minimal couplings introduced by this gauging  break supersymmetry; however, it is restored by introducing  additional terms  into the Lagrangian. 
For example, one can gauge the $SO(8)$ subgroup of the 
full $U$-duality group $E_{7(7)}$ of maximal supergravity in $D=4$~\cite{Cremmer:1978ds}, thereby turning all vector fields into
non-abelian gauge fields of $SO(8)$~\cite{de Wit:1982ig}. The gauging-induced potential does not vanish for the maximally 
($\cN=8$) supersymmetric ground state; instead, it produces a stable anti-de Sitter (AdS) vacuum. This means that, in this 
maximally supersymmetric background, all the fields of $SO(8)$ gauged $\cN=8$  supergravity form an irreducible 
supermultiplet of the AdS superalgebra $OSp(8|4,\mathbb{R})$ with even subalgebra $SO(8) \oplus Sp(4,\mathbb{R})$.  

In five dimensions, the maximal Poincar\'{e} supergravity theory has $27$ vector fields~\cite{Cremmer:1980gs}; however, 
there is no simple group of dimension $27$. The problem of gauging the maximal supergravity theory was unresolved until the massless supermultiplets of the $\cN=8$ supersymmetry 
algebra $SU(2,2|4)$ in AdS$_5$ were constructed in ref.~\cite{Gunaydin:1984fk}.  The massless five-dimensional $\cN=8$ AdS graviton supermultiplet turned out to have only  $15$ vector fields; the remaining $12$ vector fields must instead be dualized to tensor fields. Hence it was pointed out in ref.~\cite{Gunaydin:1984fk} that one can at most gauge a $SU(4)$ subgroup of the $R$-symmetry group $USp(8)$ in five dimensions. 
Gauged maximal supergravity in $D=5$ with simple gauge groups $SO(6-p,p)$ was subsequently constructed in 
refs.~\cite{Gunaydin:1984qu,Gunaydin:1985cu}.\footnote{Compact $SO(6)$ gauged supergravity was also constructed in ref.~\cite{Pernici:1985ju}.} 
In these $SO(6-p,p)$ gaugings of maximal supergravity all the supersymmetric vacua turned out to be  AdS. Shortly thereafter an $SU(3,1)$ gauged version of
maximal supergravity in five dimensions, which admits an $\cN=2$ 
supersymmetric ground state with a Minkowski vacuum, was discovered \cite{Gunaydin:1985tb}. 
In this ground state of the $SU(3,1)$ gauged maximal supergravity, one has unbroken $SU(3)\times U(1)$ gauge symmetry 
and an  $SU(2)_R$ local $R$-symmetry. The $SO(8)$ gauged $\cN=8$ supergravity in $D=4$ and $SO(6)$ gauged $\cN=8 $ supergravity in $D=5$ 
describe the low-energy effective theories of $M$-theory on $AdS_4\times S^7$ and IIB superstring over $AdS_5\times S^5$,
respectively. In general compactification of  $M$-theory or superstring theories with fluxes leads to gauged supergravity theories. 
In recent years the embedding tensor formalism for constructing gauged supergravities was developed, which is especially well suited for studying  flux compactifications. For  reviews and references on the subject we refer to refs.~\cite{Grana:2005jc,Samtleben:2008pe}.

The double-copy construction of the amplitudes of ungauged maximal supergravity has not yet been consistently or comprehensively extended 
to gauged versions of the theory. 
One conceivable obstacle in this endeavor is the fact that fully-supersymmetric ground states of gauged maximal
supergravity theories are AdS and the methods  for double-copy construction to date require a flat background. 
However, the fact that the $SU(3,1)$ gauged maximal supergravity in $D=5$ admits a stable $\cN=2$ supersymmetric ground state with vanishing cosmological constant, 
does suggest that flat-space techniques for the double-copy construction can be used 
for this theory. 

Indeed, one should expect that such double-copy constructions are allowed. Quite some time ago, Bern, De Freitas and Wong~\cite{Bern:1999bx}
constructed tree-level amplitudes in Einstein gravity coupled to YM theory, in four dimensions and without supersymmetry, 
through a clever use of the KLT relations. 
They realized that color-dressed YM amplitudes could be written in circular ways as KLT 
products between color-stripped YM amplitudes and a scalar $\phi^3$ theory. The scalar theory is invariant under two Lie groups, color and flavor (see refs.~\cite{Bern:2010yg,Bern:2011ia,Du:2011js,BjerrumBohr:2012mg,Cachazo:2013iea} for more recent work on this theory). Remarkably, after applying the KLT formula 
to the color-stripped amplitudes, the global flavor group of the scalars was promoted to the gauge group of the gluons.
They then minimally coupled the  $\phi^3$ theory to YM theory, which, through KLT, allowed them to compute single-trace tree-level amplitudes 
in a Yang-Mills/Einstein theory~\cite{Bern:1999bx}, {i.e.} the amplitudes with the highest power of the trilinear scalar coupling $g'$. 
The gravitationally-dressed Parke-Taylor amplitudes 
obtained this way match those previously constructed by Selivanov in  \cite{Selivanov:1997aq, Selivanov:1997ts} through 
a  generating function method.
Unfortunately, without the modern framework of color/kinematics duality, it was not clear how to generalize this construction to subleading powers of $g'$, 
multiple-trace amplitudes, and more importantly to loop-level amplitudes. Instead, this generalization will be achieved in the current work.

Focusing on suitable theories with flat-space vacuum, it is well known that four- and five-dimensional 
$\cN\leq 4$ supergravities with additional matter multiplets admit a very rich family of gaugings that preserve supersymmetry. 
For example, the $\cN=4$ supergravity coupled to $n$ vector supermultiplets has the global symmetry 
group $SO(6,n)\times SU(1,1)$  in $D=4$ and $SO(5,n)\times SO(1,1)$ in $D=5$. The family of four- and 
five-dimensional Maxwell-Einstein supergravity theories (MESGTs) corresponding to $\cN=2$ supergravity coupled to $n$  vector multiplets is even richer. 
A general construction of $\cN=2$ MESGTs in $D=5$ 
was given in ref.~\cite{Gunaydin:1983bi} and a construction of matter-coupled $\cN=2$ supergravity theories in $D=4$ was given in 
refs.~\cite{de Wit:1983rz,de Wit:1984px}.
An important feature of $\cN=2$ MESGTs is the fact that all the bosonic fields are $R$-symmetry singlets. As a 
consequence,  gauging a subgroup of their global symmetry groups in $D=5$ does not introduce a potential, and 
hence one always has supersymmetric Minkowski vacua in the resulting Yang-Mills/Einstein supergravity theories (YMESGTs)
in both $D=5$ and $D=4$. Gaugings of five-dimensional $\cN=2$ MESGTs were thoroughly studied in refs.~\cite{Gunaydin:1984nt,Gunaydin:1984pf,Gunaydin:1984ak,Gunaydin:1999zx}. The gaugings of four-dimensional MESGTs were originally studied in refs.~\cite{de Wit:1984pk,Cremmer:1984hj}. For a  complete list of references, we refer the reader to the book by Freedman and van Proeyen \cite{supergravity}.

Ungauged $\cN=2$ MESGTs coupled to  hypermultiplets in four and five dimensions 
arise as low-energy effective theories of type-II superstring and $M$-theory compactified over 
a Calabi-Yau threefold, respectively.  
They can also be obtained as low-energy effective theories of the heterotic string compactified on a $K3$ surface down 
to six dimensions, followed by toroidal compactifications to five and four dimensions. In general,  six-dimensional ${\cal N}=(1,0)$ supergravity coupled to $n_T$  self-dual tensor multiplets and $n_V$ vector multiplets reduces to a $D=5$ MESGT with $(n_T +n_V +1) $ vector multiplets. Such six-dimensional supergravity theories, coupled in general to hypermultiplets,   can be obtained from $M$-theory compactified to six dimensions on 
$K3\times S^1/\mathbb{Z}_2 $ \cite{Sen:1996tz,Duff:1996rs}. They can also be obtained from $F$-theory on an elliptically fibered  Calabi-Yau threefold \cite{Ferrara:1996wv}.\footnote{For 
a review and further references on the subject of M/superstring-theory origins of $\cN=2$ supergravities in $D=5$ and $D=4$ as well as $D=6$ see ref.~\cite{Aspinwall:2000fd}.} 
If we restrict the low-energy effective theory to its vector sector, the resulting MESGT 
Lagrangian in five dimensions can be fully constructed by knowing a particular set of trilinear vector couplings. These are completely specified by a symmetric tensor $C_{IJK}$, where the indices $I,J,K$ label all the vectors in the theory including the graviphoton \cite{Gunaydin:1983bi}.    
For the MESGT sector of the $\cN=2$ supergravity theory
obtained by compactification on a Calabi-Yau threefold, the $C$ tensor simply corresponds to the triple 
intersection numbers, which are topological invariants. Whenever the $C$ tensor is invariant under some symmetry transformation, the $D=5$ MESGT Lagrangian will posses  a corresponding (global) symmetry. 
Indeed, this is a consequence of the fact that $D=5$ MESGT theories are uniquely defined by their $C$ tensors \cite{Gunaydin:1983bi}.\footnote{The only physical requirement is that the $C$ tensor must lead to positive definite kinetic-energy terms for the vector and scalar fields.} 

The special cases in which the five-dimensional $\cN=2$ MESGT has a symmetric target space have long been known in the literature \cite{Gunaydin:1983bi,Gunaydin:1986fg,deWit:1992cr,deWit:1992wf}. The MESGTs with symmetric scalar manifolds (target spaces) $G/K$ such that their $C$ tensors are $G$-invariant are in one-to-one correspondence  with Euclidean Jordan algebras of degree three, whose norm forms are given by the $C$ tensor. 
There exist an infinite family of reducible Jordan algebras $\mathbb{R}\oplus \Gamma_{ n}$ of degree three,
which describes the coupling of $\cN=2$ supergravity to an arbitrary number ($ n \geq 1$) of vector multiplets. This class of theories is named 
\emph{the generic Jordan family} in the literature, and yields target spaces of the form 
\be {SO( n -1,1)\over SO( n-1) } \times SO(1,1) \ . \ee
Additionally, there exist four unified \emph{magical} MESGTs constructed from the four simple Jordan algebras of degree three, which can be realized 
as $3\times 3$ hermitian matrices over the four division algebras $\mathbb{R}, \mathbb{C}, \mathbb{H}, \mathbb{O}$. These theories 
describe the coupling of five-dimensional $\cN=2$ supergravity to $5,8,14$ and $26$ vector multiplets, respectively.

As a first step in the systematic study of amplitudes in gauged 
supergravity 
theories, in this paper we initiate
the study of amplitudes of $\cN=2$ YMESGTs constructed as double copies using color/kinematics duality. In particular, we focus on those $\cN=2$ YMESGTs whose gauging corresponds to a subgroup of the global symmetry group of the generic Jordan family of MESGTs in $D=5$ and $D=4$ Minkowski spacetime.
Furthermore, we restrict our study to compact gauge groups, and leave to future work the study of more general gaugings in generic Jordan family as well 
as in magical supergravity theories.
We note that the gauging of the generic Jordan family of supergravities is quite important from a string-theory perspective. 
The low-energy effective theory that arises from the heterotic string compactified over $K3$ is described by six-dimensional ${\cal N}= (1,0)$ 
supergravity coupled to one self-dual tensor multiplet together with some sYM multiplets and hypermultiplets. Under dimensional reduction this theory yields 
$D=5$ YMESGTs belonging to the generic Jordan family with a compact gauge group coupled to certain number of hypermultiplets. 
We will not discuss the interactions of hypermultiplets 
in the current work; since they always appear in pairs in the effective action, they can be consistently 
truncated away.

Our main results are: (1) the construction of the ungauged amplitudes as double copies between elements from a pure $\cN=2$ sYM theory and 
a family of $\cN=0$ YM theories that can be viewed as the dimensional reductions of $D=n+4$ pure YM theories; (2) the introduction of relevant cubic scalar couplings to the $\cN=0$ gauge theory,
which, through the double copy, are responsible for the interactions of the non-abelian gauge fields of the YMESGT. 
As in ref.~\cite{Bern:1999bx}, the gauge symmetry in the supergravity theory originates from a global symmetry in the $\cN=0$ YM theory employed in the double copy.
Our construction is expected to give the complete perturbative expansion of the $S$ matrix in these theories, including the full power series in the gauge coupling and 
all multi-trace terms, at arbitrary loop 
orders.\footnote{Modulo issues that are common to all formalisms:  possible UV divergences and quantum anomalies.}
Although the current work is limited to $\cN=2$ supergravity theories, we expect that our construction straightforwardly extends 
to $\cN=4$ supergravity coupled to vector multiplets as well as to some matter-coupled supergravities with $\cN<2$. The former theory can be obtained by promoting the $\cN=2$ sYM theory to $\cN=4$ sYM, and the latter theories can be obtained by truncating the spectrum of the $\cN=2$ sYM theory, while in both cases leaving the bosonic YM theory unaltered.

In section \ref{general constraints double-copy}, we provide a brief review of color/kinematics duality, identify the types of 
gauged supergravities that can most straightforwardly be made consistent with the double-copy construction and show that, 
on dimensional grounds, gauge interactions in the supergravity theory require the introduction of cubic scalar couplings in 
one of the gauge factors.

In section \ref{Lagrangians5}, we review the Lagrangians for general MESGTs and YMESGTs in five and four dimensions, 
giving particular attention to gauged and ungauged theories belonging to the generic Jordan family. 
We show how the full Lagrangian can be constructed from the $C$ tensors controlling the $F \wedge F \wedge A$ interactions 
in five dimensions. Moreover, we discuss the fundamental role played by duality transformations 
in four dimensions and the corresponding symplectic structure. We also show that, for theories belonging to the generic 
Jordan family, it is possible to find a symplectic frame for which 
(1) the linearized supersymmetry transformations act diagonally on the flavor indices of the fields when the Lagrangian is expanded around a base point, 
and (2) the three-point amplitudes have manifest $SO( n)$ symmetry.  

Section \ref{CKsec} discusses in detail the gauge-theory factors entering the double-copy construction for theories in the generic Jordan family. 
We show that we can take as one of the factor the dimensional reduction to four dimensions of the pure YM theory in $n +4$ dimensions. 
We also identify particular cubic couplings that are responsible for the non-abelian interactions in the YMESGT obtained with the double-copy. 
These couplings have the mass dimension required by the general argument presented in section \ref{general constraints double-copy}. 

In section \ref{explicit}, we compare the amplitudes from the double-copy construction with the ones from the Lagrangian discussed in section \ref{Lagrangians5}. 
Introducing particular constrained on-shell superfields, we obtain compact expressions for the superamplitudes of the theory. 
We show that, when we employ the Feynman rules obtained from the Lagrangian in the particular symplectic frame identified at the end of section \ref{Lagrangians5}, 
the two computations lead to the same three-point amplitudes and that the double-copy superfields are 
mapped trivially ({i.e.} by the identity map) into the Lagrangian on-shell superfields. 
We argue that, since the $C$ tensors are fixed by the three-point interactions, the double-copy 
construction should continue to yield the correct amplitudes at higher points. 
The fact that the double-copy construction naturally extends to five
  dimensions plays a critical role in the argument.
We run some further 
checks on the amplitudes at four points and present compact expressions for the five-point amplitudes. 

In section \ref{oneloop} we present some amplitudes at one loop, 
while our concluding remarks  are collected in section \ref{conclusions}. 

For readers' convenience, in  appendix~\ref{appendixA} we list our conventions and provide a summary of the notation 
employed throughout the paper.
Finally, in appendix~\ref{appB} we present expansions for the various quantities entering the bosonic 
Lagrangian in the symplectic frame discussed in section \ref{Lagrangians5}. 
These expansions can be used to obtain the Feynman rules for the computation outlined in section \ref{explicit}.

\section{Designer gauge and supergravity theories \label{general constraints double-copy}}

\subsection{Color/kinematics duality: a brief review}

It has long been known \cite{KLT, KLT2} 
that, at tree level, the scattering amplitudes of gravity and 
supergravity theories related to string theory compactifications on tori exhibit a double-copy structure, being expressible as sums 
of products of amplitudes of certain gauge theories.
This structure was more recently clarified in refs.~\cite{BCJ, BCJLoop}, where it was realized that there exist underlying kinematical Lie-algebraic relations 
that control the double-copy factorization. The integrands of gauge-theory amplitudes are best arranged in a cubic (trivalent) graph-based presentation 
that exhibits a particular duality between their color and kinematic numerator factors. Once such a presentation is obtained, the
double-copy relation between the integrands of gauge-theory and gravity amplitudes extents smoothly to loop level. 

In this organization, the gauge-theory $L$-loop $m$-point amplitude for adjoint particles is given by
\begin{equation}
{\cal A}^{L-\text{loop}}_m\ =\ 
i^L \, g^{m-2 +2L } \,
\sum_{i\in \Gamma}{\int{\prod_{l = 1}^L \frac{d^D p_l}{(2 \pi)^D}
\frac{1}{S_i} \frac {n_i c_i}{\prod_{\alpha_i}{p^2_{\alpha_i}}}}} \  .
\label{LoopGauge} 
\end{equation}
Here the sum runs over the complete set $\Gamma$ of $L$-loop $m$-point graphs with only cubic vertices, including all 
permutations of external legs, the integration is over the $L$ independent loop momenta $p_l$  and the denominator is given 
by the product of  all propagators of the corresponding graph.  The coefficients $c_i$ are the color factors obtained by 
assigning to each trivalent vertex in a graph a factor of the gauge  group structure constant 
${\tilde \rf}^{\hat a \hat b \hat c} = i \sqrt{2} \rf^{\hat a \hat b \hat c }=\Tr([T^{\hat a},T^{\hat b}]T^{\hat c})$ 
while respecting the cyclic ordering of edges at the vertex.
The hermitian generators $T^{\hat a}$ of the gauge group are normalized so 
that $\Tr(T^{\hat a} T^{\hat b}) = \delta^{{\hat  a} {\hat  b}}$.
The coefficients $n_i$ are kinematic numerator factors depending on momenta, polarization vectors and spinors. For 
supersymmetric amplitudes in an on-shell superspace they will also contain Grassmann parameters representing the 
odd superspace directions. These Grassmann parameters transform in the fundamental representation of the on-shell 
$R$-symmetry group of the theory. There is one such parameter for each external state; in four-dimensional theories  
there is a close relation between the number of Grassmann parameters and the helicity of the corresponding external state.
The symmetry factors $S_i$ of each graph remove any overcount introduced by summation over all permutations of external 
legs (included in the  definition of the set $\Gamma$), as well as any internal automorphisms of the graph ({i.e.} symmetries 
of the graph with fixed external legs).  

A gauge-theory amplitude organized as in equation~\eqref{LoopGauge} is said to manifestly exhibit color/kinematics 
duality~\cite{BCJ, BCJLoop} if the kinematic numerator factors $n_i$ of the amplitude are antisymmetric at each vertex, and satisfy Jacobi relations around each propagator in one-to-one correspondence with the color factors.  Schematically, the latter constraint is
\begin{equation}
c_i + c_j + c_k =0 \qquad  \Leftrightarrow \qquad  n_i + n_j + n_k =0 \, .
\label{BCJDuality}
\end{equation} 
It was conjectured in refs.~\cite{BCJ, BCJLoop} that such a representation exists to all loop orders and 
multiplicities for wide classes of gauge 
theories.\footnote{Constraints following from the requirement that the theory be renormalizable in four dimensions have been 
discussed in ref.~\cite{Chiodaroli:2013upa}.}

While amplitudes with manifest color/kinematics duality have been explicitly constructed both at tree-level and at loop-level
in various theories \cite{KiermaierTalk, BjerrumBohr:2010hn, Mafra:2011kj,
Tye:2010dd, Monteiro:2011pc,Broedel:2011pd, Johansson:2013nsa,Litsey:2013jfa,Fu:2014pya,Naculich:2014rta,
Bern:2010tq,Carrasco:2011mn,Bern:2012uf,Carrasco:2012ca, Bern:2013yya, Nohle:2013bfa, Bjerrum-Bohr:2013iza,Ochirov:2013xba,Chiodaroli:2013upa,Johansson:2014zca}, 
they are often somewhat difficult to find because of the 
non-uniqueness of the numerators. At tree-level it is possible to test whether such a representation exists by verifying 
relations between color-stripped partial amplitudes that follow from the duality~\cite{BCJ}. Examples of such BCJ 
relations for adjoint-representation color-ordered tree amplitudes are
\bea
s_{24}A_4^{(0)}(1,2, 4, 3) &=& A_4^{(0)}(1, 2, 3, 4) s_{14}   \,,
\nn\\[2pt]
s_{24}A_5^{(0)}(1,2, 4, 3,5) &=& 
{ A_5^{(0)}(1,2, 3, 4,5) (s_{14} + s_{45})+
  A_5^{(0)}(1, 2, 3, 5, 4) s_{14}}\,,\phantom{xxxxxxxx}
\nn\\[2pt]
s_{24}A_6^{(0)}(1, 2, 4, 3, 5, 6) &=& 
{A_6^{(0)}(1, 2, 3, 4, 5, 6) (s_{14} + s_{46}+s_{45}) } \nn\\[2pt] 
&& \null + 
{A_6^{(0)}(1, 2, 3, 5, 4, 6) (s_{14} + s_{46})  }
+{A_6^{(0)}(1, 2, 3, 5, 6, 4) s_{14}  }\,.
\label{amplituderelations}
\eea
We refer the reader to ref.~\cite{BCJ} for a detailed description of all-multiplicity amplitude relations.
In section~\ref{CKvs} we shall use these relations to demonstrate that a certain Yang-Mills-scalar theory, which appears in our construction, 
exhibits color/kinematics duality.

Starting from two copies of gauge theories with amplitudes obeying color/kinematics duality and assuming that at least one of them does 
so manifestly, the amplitudes of the related supergravity theory are then trivially given in the same graph organization but with the 
color factors of one replaced by the numerator factors of the other: 
\begin{equation}
 {\cal M}^{L-\text{loop}}_m = i^{L+1} \, \left(\frac{\kappa}{2}\right)^{m-2+2L} \,
\sum_{i\in\Gamma} {\int{ \prod_{l = 1}^L \frac{d^D p_l}{(2 \pi)^D}
 \frac{1}{S_i}
   \frac{n_i {\tilde n}_i}{\prod_{\alpha_i}{p^2_{\alpha_i}}}}} \equiv \left(\frac{\kappa}{2}\right)^{m-2+2L}  {M}^{L-\text{loop}}_m  \,.
\hskip .7 cm 
\label{DoubleCopy}
\end{equation}
Here $\kappa$ is the gravitational coupling. In writing this expression one formally identifies 
$\SYMcoupling^2\rightarrow \kappa/2$;  additional parameters that may appear in the gauge-theory 
numerator factors are to be identified separately and on a case-by-case basis with supergravity parameters.
The Grassmann parameters that may appear in $n_i$ and/or ${\tilde n}_i$ 
 are inherited by the corresponding supergravity amplitudes; they imply a particular organization 
of the asymptotic states labeling supergravity amplitudes in multiplets of linearized supersymmetry. Since these 
linearized transformations -- given by shifts of the Grassmann parameters -- are inherited from the supersymmetry 
transformations of the gauge-theory factors, they need not be the same as the natural linearized supersymmetry 
transformations following from the supergravity Lagrangian and a nontrivial transformation may be necessary 
to align the double-copy and Lagrangian asymptotic states.\footnote{In the case of scalar fields this transformation
needs not be linear, {e.g.}  see ref.~\cite{Carrasco:2013ypa} for an example of a non-linear transformation. Supersymmetry then implies that other fields
should also be transformed nonlinearly, {e.g.} fields in nontrivial representations of the Lorentz group may be 
dressed with functions of the scalar fields as it is natural from a Lagrangian standpoint, see section~\ref{Lagrangians5}.
}

\subsection{Minimal couplings and the double-copy construction \label{minimalsec}}

It is interesting to explore on general grounds what types of gauged supergravity theories are consistent 
with a double-copy structure with the two factors being local field theories. We discuss here the constraints 
related to the kinematical structure of amplitudes; further constraints related to the field content of the theory 
may be studied on a case-by-case basis.
To this end let us consider a four-dimensional supergravity theory coupled to non-abelian gauge fields, which has 
a supersymmetric Minkowski vacuum. While the matter interactions depend on the details of the theory, the gravitational
interactions of matter fields and those related to them by linearized supersymmetry are universal, being determined by 
diffeomorphism invariance. Similarly, the terms linear in the non-abelian gauge fields are also universal, being determined by 
gauge invariance.  

For gauged supergravities that have a string-theory origin one may expect that a KLT-like construction \cite{KLT, KLT2}  should correctly 
yield their scattering amplitudes. Similarly, it is natural to expect that a double-copy-like construction  \cite{BCJ} would 
then extend this result to loop-level \cite{BCJLoop}. In either construction, the three-point supergravity amplitudes are simply products of three-point amplitudes of the two gauge-theory factors. Assuming that this property should hold generally, the dimensionality of these amplitudes imposes strong constraints in the structure the two gauge-theory factors. Let us look first at the interactions 
of scalars and spin-$1/2$ fermions. Since they are minimally coupled to non-abelian vector fields,
\be
{\cal L}^\text{gSG}_\phi \sim |D_\mu \phi|^2 + {\bar \zeta}D_\mu {\bar\sigma}^\mu \zeta \dots \  ,
\label{generalL3}
\ee
the scalar-vector and fermion-vector three-point amplitudes $M_3^{(0)}(\phi, \phi, A) $ and $M_3^{(0)}(\zeta, \zeta, A) $
have unit mass dimension. 

This is the same dimension as that of three-point amplitudes in a classically scale-invariant gauge theory in $D=4$,\footnote{The analysis of this subsection can be repeated in $D$ dimensions, leading to the same conclusions.}
\be
[{A}^{(0)}_3] = 1 \ ,
\ee
because 
the relevant interactions are either uniquely determined by gauge invariance to be the same as \eqref{generalL3}
or are constrained by scale invariance. 
Multiplying two such amplitudes as suggested by the expected KLT/double-copy construction leads to a
gravity amplitude of dimension
\be
[M_3^{(0)}] = 2 \ ,
\ee
where we have stripped off the dimensionful gravitational coupling.
Thus, the double-copy of classically scale-invariant gauge theories cannot lead to standard non-abelian gauge 
field interactions. To obtain a supergravity amplitude of unit dimension, classical scale invariance must be broken in one of the 
two gauge-theory factors to allow for the existence of amplitudes of vanishing mass dimension. Using the 
fact that supergravity vectors, spin-$1/2$ fermions and scalars are written in terms of gauge theory fields as
\be
V_\pm = A_\pm\otimes \phi 
\qquad
\zeta_\pm  = \lambda_\pm  \otimes \phi 
\qquad
S = \phi \otimes \phi \ ,
\ee
it is easy to see that the constraints imposed by the dimensionality of amplitudes can be satisfied if 
one of the gauge-theory factors contains a cubic scalar coupling. The dimensionful parameter $g'$ governing such a coupling
should then be related to the product of the gravitational coupling and the coupling of the supergravity gauge group, 
$g'\sim \kappa g$.


The same analysis implies that the supergravity fields that are identified with products of gauge-theory fields with nonzero helicity,
\be
\psi_-  = A_-\otimes \lambda_+
\qquad
\phi_1 = A_-\otimes A_+
\qquad
\phi_2 = \lambda_-\otimes \lambda_+ \ ,
\label{spin-spin}
\ee
cannot couple directly to the non-abelian vector potential but only to its field strength. 
It is indeed not difficult to see that any three-point amplitude, in a conventional gauge theory, with at least one field with nonzero spin has unit dimension.  Thus any supergravity amplitude obtained from such 
a double copy with at least one field of the type \eqref{spin-spin} has dimension 2, and therefore cannot be given by a minimal coupling term.

It is easy to extend the analysis above to gravitino gauge interactions.\footnote{See ref.~\cite{BjerrumBohr:2010rt} for a thorough 
discussion of gravitino interactions in ungauged supergravities from the perspective of the gauge-theory factors.} Such 
interactions arise in theories in which part of the $R$-symmetry group is gauged. 
In this case the minimal 
coupling is
\be
{\cal L}_3 \sim {\bar\psi}_\mu \gamma^{\mu\nu\rho}D_\nu\psi_\rho 
\label{gravitino3pt}
\ee
and, consequently, the two-gravitini-vector amplitudes have again unit dimension. Supplementing this with the 
helicity structure following from the structure of the Lagrangian \eqref{gravitino3pt} it follows that 
\be
M^{(0)}_3(1^{\psi_+},2^{V_+}, 3^{\psi_-})\sim \frac{[12]^4}{[13][23]^2} \ .
\ee
Since
\be
\psi_+ = A_+\otimes \lambda_+
,\qquad
\psi_- = A_-\otimes \lambda_-
,\qquad
V_+ = A_+\otimes \phi \; \oplus \;  \lambda_+\otimes\lambda_+ \ ,
\ee
Lorentz invariance implies that the only way to realize this field content is by double-copying a three-vector and a two-fermion-scalar 
amplitude. The helicity structure however is incompatible with the latter amplitude originating from a standard Lorentz-invariant 
Yukawa interaction, which 
requires that the two fermions have the same helicity. Together with the fact that such a construction leads to an amplitude of 
dimension two and the fact that the three-point vector amplitudes are uniquely determined by gauge invariance, this implies that one 
of the two gauge-theory factors must have a non-conventional dimension 3 fermion-scalar interaction. 
We leave the identification of such possible interactions for the future and instead focus in the remainder of 
this paper on trilinear scalar deformations and the corresponding YMESGTs.


\section{$\cN=2$ supergravity in four and five dimensions \label{Lagrangians5}}

\subsection{General five-dimensional Maxwell-Einstein Lagrangian \label{sec5dMESGT}}

In the present section we summarize the relevant results on MESGTs in five dimensions  following refs.~\cite{Gunaydin:1984qu,Gunaydin:1985cu,Gunaydin:1983bi,Gunaydin:1984nt,Gunaydin:1984pf, Gunaydin:1984ak,
Gunaydin:1999zx, Gunaydin:1986fg}
and adopt the conventions therein. We use a five-dimensional 
spacetime metric of ``mostly plus" signature $(- + + + +)$ 
and impose a symplectic-Majorana condition of the form 
\be \chi^{\vphantom{t} \hat \imath} = {\cal C} (\bar \chi^t)^{\hat \imath} \ee
on all fermionic quantities where $\hat \imath, \hat \jmath,..=1,2$. ${\cal C}$ is the charge conjugation matrix and the Dirac conjugate is defined 
as $\bar \chi^{\hat \imath} = \chi_{\hat \imath}^\dagger \, \Gamma^0$. 
The field content of the general five-dimensional MESGT with $\tilde n$ vector multiplets is 
\be \Big\{ e^m_\mu , \Psi^{\hat \imath}_\mu, A^I_\mu, \lambda^{\hat \imath a}, \varphi^x \Big\} \ , \ee
where the indices $I,J=0,1,\ldots, \tilde n$ label the vector fields. The scalar fields $\varphi^x$ can be thought of as coordinates of a
$\tilde n$-dimensional target-space manifold ${\cal M}$; $a,b= 1,\ldots, \tilde n$ and $x,y= 1,\ldots,\tilde n$ are flat and curved 
indices in the target-space manifold, respectively. 
$R$-symmetry $SU(2)_R$ indices $\hat \imath, \hat \jmath,..$  are raised and lowered by the symplectic metric 
$\epsilon_{\hat \imath \hat \jmath}$ and its inverse $\epsilon^{\hat \imath \hat \jmath}$. 
Since our analysis focuses on amplitudes involving bosons, we write explicitly here 
only the  bosonic part of the Lagrangian, which takes on the form, 
\bea
e^{-1} {\cal L} &=& -{R \over 2} -{1\over 4} \text{\textit{\aa}}_{IJ} F^I_{\mu \nu} F^{J \mu \nu }   
- {1 \over 2} g_{xy} \partial_\mu \varphi^x \partial^\mu \varphi^y+{e^{-1} \over 6 \sqrt{6} } C_{IJK} \epsilon^{\mu \nu \rho \sigma \lambda} F^{I}_{\mu \nu}
F^J_{\rho \sigma} A^K_{\lambda} \ , \label{ungaugedL} \eea
where $g_{xy}$ is the metric of the $\tilde n$-dimensional target-space manifold ${\cal M}$.
The theory is uniquely specified by the symmetric constant tensor $C_{IJK}$ that appears in the $F\wedge F \wedge A$ term. The only constraint on the $C$ tensor is the physical 
requirement that the kinetic-energy terms of all fields be positive-definite. The connection between the $C$ tensor and the 
``field-space metrics" of the kinetic-energy terms of the vector and scalar fields proceeds via the associated cubic prepotential,
\be
N(\xi) = \Big( {2 \over 3}\Big)^{3/2} C_{IJK} \xi^I \xi^J \xi^K \ . 
\ee
The scalar manifold ${\cal M}$  can always be  interpreted as the codimension-1 hypersurface with equation $N(\xi)=1$ 
in the $\tilde n+1$-dimensional ambient space $\mathcal{C}$ with coordinates $\xi^I$ and endowed with the metric
\be 
a_{IJ} = -{1 \over 2} \partial_I \partial_J \ln N (\xi) \ . 
\ee
The metric of the kinetic energy  term for the vector fields is the restriction of the ambient-space metric $a_{IJ}$ 
to the hypersurface ${\cal M}$, while
the target space metric $g_{xy}$ is the induced metric on ${\cal M}$,
\bea 
\text{ \textit{\aa}}_{IJ}(\varphi) = a_{IJ} \big|_{N(\xi)=1} 
\ , \qquad  \qquad 
g_{xy}(\varphi) = \text{ \textit{\aa}}_{IJ} \partial_x \xi^I \partial_y \xi^J \ . 
\eea
The vielbeine $(h_I, h^a_I)$ and their inverses $(h^I, h^I_a)$ for the metric $\text{\textit{\aa}}_{IJ}$ obey the standard algebraic relations,
\bea 
\text{\textit{\aa}}_{IJ}= h_I h_J + h^a_I h^a_J \ , & \qquad & \no 
h^I_a  h^J_b \text{\textit{\aa}}_{IJ} = \delta_{ab} \ , \\  
h^I h^J \text{\textit{\aa}}_{IJ} = 1  \ , & \qquad &
h^I h_{Ia} = h_I h^{Ia}=0 \ .
\eea
They may be expressed in terms of the (derivatives of the) prepotential and the embedding coordinate $\xi$ and their derivatives as
\bea 
h^I(\varphi) &=& \sqrt{ 2 \over 3} \xi^I \big|_{N=1} \ , \\ 
h_I(\varphi) &=& \sqrt{ 1 \over 6} \partial_I \ln N (\xi) \big|_{N=1} \ , \\
h_{x}^I(\varphi) &=& - \sqrt{2 \over 3} \partial_x h^I \ , \\
h_{Ix} (\varphi) &=& \sqrt{2 \over 3} \partial_x h_I 
\ .
\eea
Last but not least, the kinetic-energy term of the scalar fields can also be expressed in terms of the $C$ tensor as
\bea
-\frac{1}{2} g_{xy} (\partial_\mu \phi^x)(\partial^\mu \phi^y) = \frac{3}{2} C_{IJK} h^I (\partial_\mu h^J) (\partial^\mu h^K ) \ .
\eea

The supersymmetry transformation laws of MESGTs are (to leading order in fermion fields)
\begin{eqnarray}\label{trafo2}
\delta e_{\mu}^{m}&=& \frac{1}{2}{\bar{\varepsilon}}^{\hat \imath}
\Gamma^{m}\Psi_{\mu \hat \imath}\nonumber \ , \cr
\delta \Psi_{\mu}^{\hat \imath} &=&\nabla_{\mu}\varepsilon^{\hat \imath}+\frac{i
}
{4\sqrt{6}}h_{\ti}(\Gamma_{\mu}^{\:\:\:\nu\rho}-4\delta_{\mu}^{\nu}
\Gamma^{\rho}) F_{\nu\rho}^{\ti}\varepsilon^{\hat \imath} \ , \cr
\delta A_{\mu}^{I}&=& -\frac{1}{2}h_{\ta}^{\ti}{\bar{\varepsilon}}^{\hat \imath}
\Gamma_{\mu}\lambda_{\hat \imath}^{\ta}+\frac{i\sqrt{6}}{4}h^{\ti} 
{\bar{\Psi}}_{\mu}^{\hat \imath}\varepsilon_{\hat \imath}\nonumber \ , \cr
\delta \lambda^{\hat \imath\ta}  &=& -\frac{i}{2}f_{\tx}^{\ta}
\Gamma^{\mu}(\partial_{\mu}
\varphi^{\tx})\varepsilon^{\hat \imath} + \frac{1}{4}h_{\ti}^{\ta}
\Gamma^{\mu\nu}
\varepsilon^{\hat \imath} F_{\mu\nu}^{\ti} \ , \\
\delta \varphi^{\tx}&=&\frac{i}{2}f^{\tx}_{\ta}{\bar{\varepsilon}}^{\hat \imath}
\lambda_{\hat \imath}^{\ta} \ ,
\end{eqnarray}
where $f^{\tx}_{\ta}$ denotes the $\tilde n$-bein on the target space. From the above transformation laws, one  can identify the field strength of the ``physical" (dressed) graviphoton to be $ h_{\ti} F_{\nu\rho}^{\ti}$ 
and the linear combinations of the vector field strengths that are superpartners of dressed spin-${1/2}$ fields $\lambda^{a \hat \imath}$, 
given by $h_{\ti}^{\ta} F_{\mu\nu}^{\ti}$.
The requirement that the $C$ tensor must lead to positive-definite kinetic-energy terms implies that at a certain point in the 
ambient space $\mathcal{C}$ the metric $\text{\textit{\aa}}_{IJ}$  can be brought to the diagonal form 
 $\text{\textit{\aa}}_{IJ}|_c= \delta_{IJ}$ by a coordinate transformation. This point is referred to as the base point $c^I$. 
 The base point $c^I$ lies in the domain of positivity with respect to the metric $a_{IJ}$ of the ambient space $\mathcal{C}$. 
Choosing the base point as 
\be 
c^I = \Big( \sqrt{3 \over 2}, 0, \ldots, 0 \Big) 
\ee
one finds that the most general $C$ tensor compatible with positivity can be brought to the  form
 \be 
 C_{000} = 1 , \qquad  C_{0ab}= -{1 \over 2} \delta_{ab} , \qquad C_{00a} = 0,  \qquad  a,b,c=1,\ldots, \tilde n  \ ,  \label{Ccan} 
 \ee
with the remaining components  $ C_{abc}$  being completely arbitrary.

The global symmetry group of the MESGT is simply given by the invariance group of the tensor  $C_{IJK}$. These global symmetry transformations correspond to isometries of 
the target space ${\cal M}$.  The converse is however not true. There exist theories in which 
not all the isometries of the target manifold extend to global symmetries of the full $\cN=2$ MESGT, 
e.g. the generic non-Jordan family to be discussed  below. 

Since our main goal is to understand the double-copy structure of gauged or ungauged YMESGTs, 
we will consider only theories whose $C$ tensors  admit symmetries, in particular, those theories whose scalar manifolds are symmetric spaces  
$\mathcal{M}= G/H$ such that $G$ is a symmetry of the  $C$ tensor.\footnote{In addition 
to the MESGTs defined by Euclidean Jordan algebras of degree three, there exists an infinite family of theories whose scalar manifolds are
symmetric spaces of the form
 \bea 
 {\cal M} = {SO(1, \tilde n) \over SO( \tilde n)} \ , \qquad \tilde n>1 \ . 
 \nonumber
 \eea
This infinite set of MESGTs is called the generic non-Jordan family. The
isometry group $SO(1,\tilde n)$ of the scalar manifold of the generic non-Jordan
family is broken by supergravity couplings down to the semi-direct product subgroup 
$ [SO(1,1)\times SO(\tilde{n}-1)]\ltimes T_{(\tilde{n}-1)} $ \cite{deWit:1992cr}.}
In the latter theories, 
the cubic form defined by the $C$ tensor can be identified with the norm  form of a Euclidean Jordan algebra of degree three.  
There are four simple Jordan algebras of degree three denoted  as $J_3^{\mathbb{A}}$, which can be realized as hermitian $3\times 3$ matrices over the four  division algebras $\mathbb{A}= \mathbb{R},\mathbb{C},\mathbb{H},\mathbb{O}$. The scalar manifolds  $\mathcal{M}_5(J^{\mathbb{A}})$ of MESGTs defined by the simple Jordan algebras in five dimensions are as follows:
\bea \mathcal{M}_5(J_3^{\mathbb{R}})= \frac{SL(3,\mathbb{R})}{ SO(3)}, & \qquad &  \mathcal{M}_5(J_3^{\mathbb{H}})=\frac{SU^*(6)}{USp(6)},\nn \\ 
\mathcal{M}_5(J_3^{\mathbb{C}}) =\frac{SL(3,\mathbb{C})}{ SU(3)},  & \qquad &  \mathcal{M}_5(J_3^{\mathbb{O}})= \frac{E_{6(-26)}}{ F_4} .  \eea
They describe the coupling of 5, 8, 14 and 26 vector multiplets to $\cN=2$ supergravity and are referred to as ``Magical supergravity theories". Their global symmetry groups $G$ are   simple  and all the vector fields including the graviphoton transform in a single irreducible representation under $G$. Hence they are unified theories. 
The quaternionic magical theory with the global symmetry group $SU^*(6)$ can be obtained by a maximal truncation of 
the $\cN=8$ supergravity in five dimensions. 

There also exists an infinite family of reducible Jordan algebras $\mathbb{R}\oplus \Gamma_{\tilde{n}}$  of degree three 
whose cubic norms factorize as a product of linear and quadratic form. 
The MESGTs defined by them are referred to as the generic Jordan family; their scalar manifolds are of the form
 \be {\cal M}_5(\mathbb{R}\oplus \Gamma_{\tilde{n}}) = {SO(\tilde n-1,1)\times SO(1,1) \over SO(\tilde n-1)} \ , \qquad \tilde n \geq 1 . \ee
and describe the  coupling of $\tilde n$ vector multiplets to ${\cal N}=2$ supergravity.
 
 The magical supergravity theories can be truncated to MESGTs belonging to the generic Jordan family. Below we list the target spaces of  maximal 
 truncations of this type: 
 \bea 
\mathcal{M}_5(J_3^{\mathbb{R}})= \frac{SL(3,\mathbb{R})}{ SO(3)} \longrightarrow \frac{SO(2,1)\times SO(1,1)}{SO(2)} \ , \nn
 \\ \mathcal{M}_5(J_3^{\mathbb{C}}) =\frac{SL(3,\mathbb{C})}{ SU(3)} \longrightarrow \frac{SO(3,1)\times SO(1,1)}{SO(3)} \  ,  \\
 \mathcal{M}_5(J_3^{\mathbb{H}})=\frac{SU^*(6)}{USp(6)}\longrightarrow \frac{SO(5,1)\times SO(1,1)}{SO(5)}\nn  \ , \\  \mathcal{M}_5(J_3^{\mathbb{O}})=
\frac{E_{6(-26)}}{ F_4}\longrightarrow \frac{SO(9,1)\times SO(1,1)}{SO(9)} \ . \nn \eea

We should note that the $C$ tensors of MESGTs defined by Euclidean Jordan algebras of degree three, generic Jordan as well as the  
magical theories, satisfy the adjoint identity

\begin{equation}
   C^{IJK} C_{J(MN} C_{PQ)K} = \delta^{I}_{(M} C_{NPQ)} \ ,
\end{equation}
where the indices are raised with the inverse ${\stackrel{\circ}{a}}{}^{IJ}$ of the vector field space metric ${\stackrel{\circ}{a}}{}_{IJ}$. 
Furthermore the $C$ tensor is an invariant  tensor of their global symmetry groups $G$ such that we have
\be
C_{IJK}=C^{IJK} \ .
\ee

In this paper we will be considering MESGTs belonging to  the generic Jordan family. The {\it natural basis} for defining 
the $C$ tensor for this family is given by identifying the prepotential with the cubic norm of the underlying Jordan algebra, which is
\be N(\xi) = \sqrt{2} \xi^0 \Big( (\xi^1)^2 - (\xi^2)^2 - \ldots - (\xi^{\tilde n})^2 \Big) \ . \ee
The    $C$ tensor in the natural basis is given by 
 \be C_{011} = {\sqrt{3} \over 2} \ , \qquad C_{0rs} = - {\sqrt{3} \over
2} \delta_{rs} \ , \qquad r,s=2,\ldots, \tilde n \ , \label{Creg}\ee
and the corresponding base point is at 
\be c^I = \Big( {1 \over\sqrt{ 2}}, 1, 0 , \ldots, 0 \Big) \ . \label{base-generic} \ee

\subsection{YMESGTs in five dimensions \label{LagrangianDiscussion}}

Let us now review the gaugings of  a $\cN=2$ MESGT in five dimensions whose full  symmetry group is the product of $R$-symmetry $SU(2)_R$ 
with the global invariance group $G$ of the $C$ tensor. If one gauges only a subgroup $K$ of the global symmetry $G$, 
the resulting theories are referred to as YMESGTs, 
which can be  compact or non-compact. 
If one gauges only a subgroup of $SU(2)_R$ the resulting theories  are called gauged MESGTs. 
If one gauges a subgroup of $SU(2)_R$ as well as a subgroup of $G$ simultaneously then they are called gauged  YMESGTs. 
In gauging a subgroup of the global symmetry group $G$ those non-gauge vector fields that transform non-trivially under the 
non-abelian gauge group must be dualized to  tensor fields that satisfy odd-dimensional self-duality conditions \cite{Gunaydin:1999zx}.  
We list the possibilities for gauging the generic Jordan family 
of MESGTs without tensor fields  in Table \ref{tabgaugings}. 

\begin{table}[t]
 \begin{center}
\begin{tabular}{ll}
 $K \subseteq SU(2)_R$ & Gauged Maxwell-Einstein supergravity \\
 $K \subseteq SO(\tilde n-1)$ & Compact Yang-Mills/Einstein supergravity  \\
 $K \subseteq SO(\tilde n-1,1)\times SO(1,1)$ & Non-compact Yang-Mills/Einstein supergravity  \\
 $K \subseteq SU(2)_R \times SO(\tilde n-1) $ & Compact gauged Yang-Mills/Einstein sugra  \\
 $K \subseteq SU(2)_R \times SO(\tilde n-1,1) \times SO(1,1)$ & Non-compact gauged Yang-Mills/Einstein sugra \\
\end{tabular}
\caption{List of possible gaugings of generic Jordan family of ${\cal N}=2$ MESGTs without tensor fields. \label{tabgaugings}}
\end{center}
\end{table}
In this paper 
we will focus on YMESGTs with compact gauge groups
obtained by gauging a subgroup of the global symmetry group of generic Jordan family. 
We leave the study of more general gaugings in generic Jordan as well as magical supergravity theories  in the framework provided by the double-copy 
to future work.

Consider a group $G$ of symmetry  transformations acting on the coordinates of the ambient space as 
\be \delta_{\alpha} \xi^I = (M_r)^I_{\ J} \xi^J \alpha^r \ ,   \ee
where $M_r$ are constant (anti-hermitian) matrices that satisfy the  commutation relations of $G$,
\be [ M_r, M_s ] = f_{rs}^{\ \ \ t} M_t \ . \ee
If $G$ is a symmetry of the Lagrangian of the MESGT, it must satisfy
\be 
(M_r)_{(I}^{~L} C_{JK)L} =0 \ .
\ee
The bosonic fields of the theory transform as 
\bea \delta_\alpha \varphi^x &=& K^x_r \alpha^r \ , \no \\
\delta_{\alpha} A^I_\mu & = & (M_r)^I_{\ J} A^J_\mu \alpha^r \ ,
\eea
where the field-dependent quantity $K^x_r$ is the Killing vector that generates the corresponding isometry of the scalar manifold $\mathcal{M}$. It can be expressed as  
\bea  K^x_r &=& - \sqrt{ 3 \over 2} (M_r)^J_{\ I} h_J h^{I x} .  
\eea
The scalar field dependent quantities $h^I(\varphi^x)$ transform just like the vector fields
\bea
\delta_{\alpha} h^I (\varphi^x)& = & (M_r)^I_{\ J} h^J(\varphi^x)  \alpha^r 
\eea 
Spin-$1/2$ fields transform under the maximal compact subgroup of the global symmetry group as 
\bea
\delta_\alpha \lambda^a_{\hat \imath}  =  L^{ab}_r \lambda^b_{\hat \imath} \alpha^r 
\eea
where 
\[ L^{ab}_r =  (M_r)^J_{\ I} h^{[a|}_J h^{I|b]} - \Omega^{ab}_x K^x_r \ . \]
with $\Omega^{ab}_x $ denoting the spin connection over the scalar manifold $\mathcal{M}$.
The remaining fields (gravitinos and graviton) are invariant under the action of global symmetry group $G$.

To gauge a subgroup $K$ of the global symmetry group $G$  of a ${\cal N}=2$ MESGT in five dimensions, we  
then split the vector fields as $A_\mu^I=(A^0_\mu,A^m_\mu,A^r_\mu)$ so that $A^r_\mu$ ($r,s,t= \tilde n +1 - \text{dim}\ K, \dots, \tilde n$)  
transform in the adjoint 
of a compact subgroup $K$ of $G$, and $A^0_\mu,A^m_\mu$ are $K$-singlets ($m,n,o=1,\ldots \tilde n- \text{dim} \ K$).
We shall refer to the latter as spectator fields and pick a set of matrices $M$ such that 
\be 
(M_r)^s_{\ t}=  f^{rst} \ , \qquad (M_r)^m_{\ s}= (M_r)^s_{\ m} = (M_r)^m_{\ n} = 0 \ . 
\label{group_generators} 
\ee

Compact YMESGTs obtained with this construction from the generic Jordan family of MESGTs in five dimensions 
always have at least two spectator fields -- the vector field in the $D=5$ gravity multiplet and an additional one. 
For the sake of simplicity we will not consider theories with more spectators and assume that
\be \tilde n = \text{dim} \  K + 1  \ee
from now on. 

The Lagrangian of the desired YMESGT is then obtained from (\ref{ungaugedL}) replacing
\bea  \partial_{\mu} \varphi^x &\rightarrow& \partial_{\mu} \varphi^x + g A^s_\mu K^x_s \ , \\ 
\nabla_{\mu} \lambda^{\hat \imath a} &\rightarrow& \nabla_{\mu} \lambda^{\hat \imath a} + g L^{ab}_t A^t_\mu \lambda^{\hat \imath b} \ , \\
F_{\mu \nu}^I &\rightarrow& \mathcal{F}^I_{\mu\nu}= 2 \partial_{[\mu} A^I_{\nu]} + g f^I_{\ JK} A^J_\mu A^K_\nu \ . 
\eea
with the caveat that 
the $F\wedge F \wedge A$ term for the gauge fields must be covariantized as follows,
\bea {e^{-1} \over 6\sqrt{6}} C_{IJK} \epsilon^{\mu\nu\rho\sigma\lambda}\left\{ 
 F^I_{\mu \nu} F^J_{\rho \sigma} A^K_{\lambda} + {3 \over 2} g f^{K}_{ J' K'} F^I_{\mu \nu} A^J_\rho A^{J'}_\sigma A^{K'}_\lambda +
{3\over 5} g^2 A^I_\mu f^{J}_{ I' J'} A^{I'}_\nu A^{J'}_\rho f^{K}_{K' L'} A^{K'}_\sigma A^{L'}_{\lambda} \right\} \no \\ \eea 
where we formally introduced the structure constants 
$f^I{}_{JK}$, which vanish when any one of the indices corresponds to the spectator fields. The only non-vanishing components are those of the compact gauge group $K$ , 
i.e $f^r{}_{st}$ with $r,s,t = 2,3,\ldots \tilde n$.

The bosonic sector of the resulting YMESGT is then given by
\bea
 \label{gaugedL} 
e^{-1} {\cal L} &=& -{R \over 2} -{1\over 4} \text{\textit{\aa}}_{IJ} \mathcal{F}^I_{\mu \nu} \mathcal{F}^{J \mu \nu }   
- {1 \over 2} g_{xy} \mathcal{D}_\mu \varphi^x \mathcal{D}^\mu \varphi^y +  \no 
{e^{-1} \over 6\sqrt{6}} C_{IJK} \epsilon^{\mu\nu\rho\sigma\lambda} \left\{ \vphantom{1\over 2}
 F^I_{\mu \nu} F^J_{\rho \sigma} A^K_{\lambda} \right. \\ 
&& \qquad \left. + {3 \over 2} g f^{K}{}_{ J' K'} F^I_{\mu \nu} A^J_\rho A^{J'}_\sigma A^{K'}_\lambda +
{3\over 5} g^2 A^I_\mu f^{J}{}_{ I' J'} A^{I'}_\nu A^{J'}_\rho f^{K}{}_{K' L'} A^{K'}_\sigma A^{L'}_{\lambda} \right\} \ , 
\eea

Note that the five-dimensional YMESGT Lagrangian does not  
have a potential term and hence admits Minkowski ground states. 
However, to preserve supersymmetry under gauging one introduces a Yukawa-like term involving scalar fields and spin-$1/2$ fields
\begin{equation}
\mathcal{L}'= -\frac{i}{2}g{\bar{\lambda}}^{i a}\lambda_{i}^{b}K_{r[a}
h^{r}_{b]}.
\label{gaugerelatedfermionbilinear}
\end{equation}
One may understand the need for such a term by noticing that in the limit of vanishing gravitational constant YMESGTs should become flat space non-abelian gauge theories, which necessarily exhibit Yukawa couplings.

\subsection{Four Dimensional $\cN =2$ MESGTs and YMESGTs via dimensional reduction \label{appA} \label{Lagrangians4}}

Since in later sections we will compute scattering amplitudes of YMESGTs (and MESGTs) in four dimensions, it is useful to review 
following ref.~\cite{Gunaydin:2005bf} the construction of the bosonic sector of these theories by dimensional reduction from five dimensions. 
To distinguish the  four- and five-dimensional  fields we shall denote the five-dimensional field indices with hats in this section.

For dimensional reduction we make the Ansatz for the f\"unfbein as follows
\be \hat e^{\hat m}_{\hat \mu} = \left( \begin{array}{cc} e^{- {\sigma \over 2} } e^m_\mu & 2 W_\mu e^{\sigma} \\ 
0 & e^\sigma \end{array} \right) \ , \ee
which implies $\hat e = e^{-\sigma} e$, where $e=$det$(e^m_\mu)$. We shall denote the field strength of the vector $W_\mu$ coming 
from the graviton in five dimensions as $W_{\mu \nu}$:
\be W_{\mu \nu} = 2 \partial_{[\mu} W_{\nu]} \ . \ee
The five-dimensional  vector fields $a^I_{\hat \mu}$ decompose into four-dimensional  vector fields $A^I_\mu$ and four-dimensional scalars $A^I$ as follows 
\be a^I_{ \hat \mu } = \left( \begin{array}{c}  A^I_\mu \\ A^I_5 \end{array} \right) =
\left( \begin{array}{c}  A^I_\mu + 2 W_\mu A^I \\ A^I \end{array} \right) \ . \ee
Note that the four-dimensional abelian field strengths 
$F^I_{\mu \nu}= \partial_\mu A^I_\nu - \partial_\nu A^I_\mu $ are  invariant  with respect to the $U(1)$ symmetry 
of the compactified circle. 
 The bosonic sector of dimensionally-reduced, four-dimensional action of the MESGT is then \cite{Gunaydin:2005bf}
\bea e^{-1} {\cal L}^{(4)} &=& -{1\over 2} R - {1 \over 2 } e^{3 \sigma} W_{\mu \nu} W^{\mu \nu} - {3 \over 4} \partial_{\mu} \sigma \partial^\mu \sigma \no \\
&&-{1\over 4} e^{\sigma} \text{\it \aa}_{IJ} (  F^I_{\mu \nu} + 2 W_{\mu \nu} A^I) (  F^{J\mu \nu} + 2 W^{\mu \nu} A^J) \no \\
&& -{1\over 2} e^{- 2 \sigma} \text{\it \aa}_{IJ} \partial_\mu A^I \partial^\mu A^J - 
{3\over 4}  \text{\it \aa}_{IJ} \partial_\mu h^I \partial^\mu h^J \no \\
&& + {e^{-1} \over 2 \sqrt{6}} C_{IJK} \epsilon^{\mu\nu\rho\sigma} \big\{ F^I_{\mu \nu} F^J_{\rho \sigma} A^K + 
2 F^I_{\mu \nu}W_{\rho \sigma} A^J A^K + {4 \over 3} W_{\mu \nu} W_{\rho \sigma} A^I A^J A^K  \big\}  . \qquad
\eea
The  four-dimensional scalar manifold geometry is defined  by $(\tilde n+1)$ complex 
coordinates \cite{Gunaydin:1983bi}
\be 
z^I  := {1 \over \sqrt{2}} A^I + {\sqrt{3} \over 2} i e^\sigma h^I \ .
\ee
One can write the four-dimensional Lagrangian as
\be 
e^{-1} {\cal L}= -{1\over 2} R - g_{I \bar J} \partial_\mu z^I \partial^\mu \bar z^J 
+ {1 \over 4} \Im \cN_{AB} F^A_{\mu \nu } F^{B \mu \nu} - {e^{-1} \over 8} 
\epsilon^{\mu\nu\rho\sigma} \Re \cN_{AB} F^A_{\mu \nu} F^B_{\rho \sigma} \ ,  \label{L4d}  
\ee
where $A,B= -1,0,1,\ldots,\tilde n$, $g_{I \bar J}= \text{\it \aa}_{IJ}$, and the period matrix $\cN_{AB}$ is given by
\bea 
\cN_{-1 -1} &=& - { \sqrt{2} \over 3 \sqrt{3}} C_{IJK} A^I A^J A^K - {i\over 2} \Big( e^\sigma \text{\it \aa}_{IJ} A^I A^J + {1 \over 2} e^{3 \sigma} \Big) \ , \no \\
\cN_{-1  I} &=&  { 1 \over \sqrt{3}} C_{IJK}  A^J A^K +  {i \over \sqrt{2}}  e^\sigma \text{\it \aa}_{IJ}  A^J \ ,\no \\
\cN_{IJ} &=& - {2 \sqrt{2} \over \sqrt{3}} C_{IJK}  A^K -  i e^\sigma \text{\it \aa}_{IJ}  \ .  \no 
\eea
The vector field $-2\sqrt{2} W_\mu$ is denoted as $A^{-1}_\mu$ and its field strength as
\be 
F^{-1}_{\mu \nu} := - 2\sqrt{2} W_{\mu \nu} \ . 
\ee

The scalar manifolds of magical supergravity theories defined by simple Jordan algebras of degree three in 
four dimensions are the following hermitian symmetric spaces 

\bea \mathcal{M}_4(J_3^{\mathbb{R}})= \frac{Sp(6,\mathbb{R})}{ U(3)}, &\qquad &  \mathcal{M}_4(J_3^{\mathbb{H}})=\frac{SO^*(12)}{
U(6)}, \nn \\ 
\mathcal{M}_4(J_3^{\mathbb{C}})
= \frac{SU(3,3)}{ S (U(3)\times U(3))},   &\qquad&  \mathcal{M}_4(J_3^{\mathbb{O}})= \frac{E_{7(-25)}}{ E_6\times U(1)}  \ . \eea
The scalar manifolds of generic Jordan family of MESGTs in $D=4$ are 

 \be {\cal M}_4(\mathbb{R}\oplus \Gamma_{\tilde{n}}) = {SO(\tilde n,2)\times SU(1,1) \over SO(\tilde
n)\times SO(2) \times U(1)} \ , \qquad \tilde n \geq 1 . \ee

Our focus in the paper will be mainly on gaugings of the generic Jordan family of MESGTs. Motivation for studying the generic Jordan family from a
string theory perspective derives from the fact that the vector-multiplet moduli spaces of heterotic string theories compactified on  $K3\times S^1$ are precisely 
of the generic Jordan type. In the corresponding superpotential 
\be \mathcal{F} = s ( t^1 t^1 - t^r t^r ) \ee
 the singlet modulus $s$ is simply the dilaton. The cubic superpotential  is exact in five dimensions. The dilaton factor corresponding 
 to the scale symmetry $SO(1,1)$  of the five-dimensional $U$-duality group gets extended by an extra scalar, the axion, under  dimensional reduction 
 to four dimensions and together they parametrize the $SU(1,1)/U(1)$ factor in the four-dimensional moduli space 
 $ SU(1,1)/U(1) \times SO(\tilde n,2)/SO( \tilde n)\times SO(2)$. The four-dimensional supergravity moduli space of the generic Jordan family gets corrections due to target space  
 instantons in the string theory. There is a corresponding picture in the type-IIA string due to the duality between 
 type-IIA theory on a Calabi-Yau threefold and heterotic string on $K3\times T^2$.  We refer the reader to the review \cite{Aspinwall:2000fd} for a detailed discussion of this duality and the references on the subject. 
 We should note that  non-abelian gauge interactions in lower dimensional effective theories of heterotic string theory descend, in general, directly from the non-abelian gauge symmetries in ten dimensions.  This is to be contrasted  with compactifications of $M$-theory or type-II 
 superstring theories  on Calabi-Yau manifolds without any isometries. The latter  theories can develop enhanced non-abelian symmetries at certain points in their 
 moduli spaces and  the corresponding low-energy effective theories are described by YMESGTs coupled to hypermultiplets. Detailed examples of such symmetry enhancement  
 both in five and four dimensions were studied in refs.~\cite{Mohaupt:2001be,Louis:2003gj}.

The dimensional reduction of the five-dimensional YMESGTs without tensor fields leads to the four-dimensional Lagrangian
\begin{eqnarray}
 e^{-1}\mathcal{L}^{(4)} &=&-\frac{1}{2}R -\frac{3}{4}
 {\stackrel{\circ}{a}}_{I J}(\mathcal{D}_{\mu}\tilde{h}^{I})(\mathcal{D}^{\mu}\tilde{h}^{J})
-\frac{1}{2}e^{-2\sigma}{\stackrel{\circ}{a}}_{IJ}(\mathcal{D}_{\mu} A^{I})(\mathcal{D}^{\mu}A^{J}) \nonumber\\
& &  -\frac{1}{4}e^{\sigma}{\stackrel{\circ}{a}}_{IJ}(\mathcal{F}_{\mu\nu}^{I}+2W_{\mu\nu}A^{I}
 )( \mathcal{F}^{J\mu\nu}+2W^{\mu\nu}A^{J})  -\frac{1}{2}e^{3\sigma}W_{\mu\nu}W^{\mu\nu} \nonumber \\
   & &   +\frac{e^{-1}}{2\sqrt{6}} C_{IJK}\epsilon^{\mu\nu\rho\sigma} \Big\{
\mathcal{F}_{\mu\nu}^{I}\mathcal{F}_{\rho\sigma}^{J}A^{K} + 2 \mathcal{F}_{\mu\nu}^{I}W_{\rho\sigma}A^{J}A^{K} +\frac{4}{3}W_{\mu\nu}W_{\rho\sigma} A^{I}A^{J}A^{K} \Big\} \nonumber\\
       & & -g^2 P_4  \ ,\label{redlag2}
\end{eqnarray}
where
\begin{eqnarray}
\mathcal{D}_{\mu}A^{I} & \equiv  & \partial_{\mu} A^{I} +g A_{\mu}^{J}f_{JK}^{I}A^{K}\\
\mathcal{F}_{\mu\nu}^{I}  &  \equiv  &    2\partial_{[\mu}A_{\nu]}^{I} + gf_{JK}^{I}A_{\mu}^{J}A_{\nu}^{K} \\
\mathcal{D}_{\mu}\tilde{h}^{I}& \equiv & \partial_{\mu} \tilde{h}^{I} + g A_{\mu}^{r}(M_r)^{I}{}_{K}\tilde{h}^{K} \ ,
\end{eqnarray}
and the four-dimensional scalar potential, $P_4$, is given by
\begin{equation}
P_4=\frac{3}{4} e^{-3\sigma}{\stackrel{\circ}{a}}_{IJ} 
(A^{r}(M_{r})^{I}{}_{K} h^{K})(A^{s}(M_{s})^{J}{}_{L}h^{L}) \ . 
\label{totalpot}
\end{equation}
The appearance of a nontrivial potential may be understood by recalling that in the limit of vanishing gravitational coupling a 
four-dimensional  YMESGT reduces to the dimensional reduction of a five-dimensional gauge theory and, as such, it has a quartic 
scalar coupling that is bilinear in the gauge-group structure constants.

\subsection{ Generic Jordan family of $4D$ $\cN =2$ YMESGTs \label{sec4dsymp}}

In this section we shall study in detail the symplectic formulation of the generic Jordan family of 
four-dimensional YMESGTs  defined by the cubic form \eqref{Creg}. Four dimensional $\cN=2$ supergravity theories coupled to vector and 
hypermultiplets were constructed in refs.~\cite{de Wit:1983rz,de Wit:1984px,de Wit:1984pk}, which showed that the prepotentials for $D=4$ MESGTs must 
be homogeneous functions of degree two in terms of the complex scalars. For those $\cN=2$ MESGTs originating from five dimensions  the prepotential 
is given by the $C$ tensor \cite{Gunaydin:1983bi,Cremmer:1984hj}. Later on, a symplectic covariant formulation of $D=4$ MESGTs was developed 
\cite{Craps:1997gp,Ceresole:1995jg} (also see ref.~\cite{supergravity} for a review and further references). 
Before we proceed, we recall some basic facts about choice of symplectic sections and existence of a prepotential. 
In a  symplectic formulation, the four-dimensional ungauged Lagrangian (\ref{L4d}) can be obtained from the prepotential 
\be 
F(Z^A) = - {2 \over 3 \sqrt{3}} C_{IJK} {Z^I Z^J Z^K \over Z^{-1}} \ ,
\label{vspecial} 
\ee
where $Z^{-1} \equiv Z^{A=-1}$.
One considers a holomorphic symplectic vector of the form
\be v = \left( \begin{array}{c} Z^A(z) \\ {\displaystyle\partial F  \over \displaystyle\partial Z^A}(z)\end{array} \right) \ ,\ee
where the $Z^A$ are $\tilde n+2$ arbitrary holomorphic functions of $\tilde n+1$ complex variables 
$z^I$, which need to satisfy a non-degeneracy condition. A standard choice is to set $Z^0 \equiv 1$ and to pick an appropriate set of $\tilde n +1$ linear functions
for the remaining $Z^I(z)$. Next, one  introduces  
a K\"{a}hler potential ${\cal K}(z,\bar z)$ defined by 
\be e^{-\cal K} = -i \langle v, \bar v \rangle = - i \Big( Z^A {\partial \bar F \over \partial \bar Z^A} - \bar Z^A {\partial  F \over \partial Z^A} 
\Big) \ .   \ee
The metric for the scalar manifold is then readily obtained as
\be g_{I \bar J} = \partial_I \partial_{\bar J} {\cal K} \ . \ee
A little more work is necessary to obtain the period matrix appearing in the 
kinetic term for the vector fields. We first introduce a second symplectic vector defined as
\be V(z,\bar z) = \left( \begin{array}{c} X^A \\  F_A   \end{array} \right) = e^{{\cal K }\over 2} v(z) \ ,  \ee
and the corresponding target-space covariant derivatives,
\bea D_{\bar I} \bar X^A &=& \partial_{\bar I} \bar X^I + {1\over 2} (\partial_{\bar I} {\cal K}) \bar X^A \ , \no \\ 
D_{\bar I} \bar F_A &=& \partial_{\bar I} \bar F_A + {1\over 2} (\partial_{\bar I} {\cal K}) \bar F_A \ . \eea
The $(\tilde n+2)\times( \tilde n+2)$ period matrix  ${\cal N}$ 
can be expressed in terms of the quantities above as
\be {\cal N}_{AB} = \big( F_A \ \ D_{\bar I} \bar F_A \big) \big( X^B \ \ D_{\bar I} \bar X^B \big)^{-1} \ .  \label{period} \ee
We should note that the set of $\tilde n+2$ holomorphic functions $Z^I$ of $\tilde n+1$ complex variables need to be chosen such that the matrix 
$\big( X^B \ \ D_{\bar I} \bar X^B \big)$ entering the expression above is invertible. 
It can be shown that with the definition (\ref{period}) the period matrix is symmetric as desired (see for example \cite{Craps:1997gp}). 

\bigskip
For the generic Jordan family with the symplectic vector that comes directly from dimensional reduction from five dimensions we have $Z^{-1}\equiv 1$ and
\bea
{\partial F \over \partial Z^{-1}} &=& {2 \over 3 \sqrt{3}} C_{IJK}Z^{I}Z^{J}Z^{K}\\
{\partial F \over \partial Z^I} &=& - {2 \over \sqrt{3}} C_{IJK}Z^{J}Z^{K}
 \label{normbasis} \ .
\eea
Only the compact subgroup $SO(\tilde n-1)$ of the full $U$-duality group $SU(1,1)\times SO(\tilde n,2)$ 
is realized linearly. One can go to a symplectic section in which the full $SO(\tilde n,2)$ symmetry is 
realized linearly. However this symplectic section does not admit a prepotential \cite{Ceresole:1995jg}.

While we will omit the fermionic part of the action as before,  the supersymmetry 
transformations of the gravitinos and spin-${1/ 2}$ fermions will be relevant. They are:
\bea
\delta e^m_{\mu} &=& {1\over 2} \bar \epsilon^{\hat \imath} \gamma^m \psi_{\mu \hat \imath} + \text{h.c.} \ , \no \\
\delta \psi^{\hat \imath}_\mu &=& D_\mu  \epsilon^{\hat \imath} + {1 \over 4} \epsilon^{\hat \imath \hat \jmath} F^{A-}_{\mu \nu} \gamma^{\mu \nu}  \Im {\cal N}_{AB} 
X^B \gamma_\mu \epsilon_{\hat \jmath} \ , \no \\
\delta A^A_{\mu} &=& {1 \over 2} \epsilon^{\hat \imath \hat \jmath} \bar \epsilon_{\hat \imath} \gamma_\mu \lambda^I_{\hat \jmath} D_I X^A 
+ \epsilon^{\hat \imath \hat \jmath}\bar \epsilon_{\hat \imath} \psi_{\mu \hat \jmath} X^A + \text{h.c.} \no \ , \\
\delta \lambda^I_{\hat \imath} &=& \gamma^\mu \nabla_\mu z^I \epsilon_{\hat \imath} - {1 \over 2} g^{I \bar J} D_{\bar J} \bar X^A \Im {\cal N}_{AB} 
F^{B-}_{\mu \nu} \gamma^{\mu \nu} \epsilon_{\hat \imath} \no \ , \\
\delta z^I &=& {1\over 2} \bar \epsilon^{\hat \imath} \lambda^I_{\hat \imath} \ .
\label{4dsusy} 
\eea
Here $F^{A\pm}_{\mu \nu}$ indicate the self-dual and anti-self-dual field strengths. With this notation 
$F^{A+}_{\mu \nu}$ and $F^{A-}_{\mu \nu}$ are complex conjugate to each other. Moreover, the dual field 
strengths are given by
\be 
G_A^{+} =  {2 i e^{-1} }  {\delta {\cal L} \over \delta F^{A+}} = {\cal N}_{AB} F^{B+} \ .
\ee
One can introduce an $Sp(2 \tilde n+4, \mathbb{R} )$ group of duality transformations acting as 
\begin{equation}
\left( \
\begin{array}{c} 
\tilde F^+ \\ \tilde G^+ 
\end{array} \right) =
\left( \begin{array}{cc} A & B \\ C & D \end{array} \right)
\left( \begin{array}{c} F^+ \\ G^+ \end{array} \right) \ ,
\end{equation}
with
\be A^t C = C^t A \ , \qquad B^t D = D^t B \ , \qquad A^tD - C^t B = {\bf 1} \ . \ee
Under such transformations the target space metric $g_{I \bar J}$ is invariant and the period matrix ${\cal N}_{AB}$ transforms as
\be \tilde {\cal N} = \big(  C + D {\cal N}\big) \big( A + B {\cal N} \big)^{-1} \ . \ee
The action is invariant under a duality transformation provided that $B=0$; transformations with $B\neq 0$ are 
non-perturbative (i.e. involve $S$-duality as seen from an higher-dimensional perspective) and are called \emph{symplectic reparameterizations}.
A duality transformation can also be enacted directly on the prepotential. 
Adopting this perspective, one needs to introduce a new prepotential
\be \tilde F = {1 \over 2} V^t(X) \left( \begin{array}{cc} C^tA & C^tB \\ D^tA & D^tB \end{array} \right) V(X) \ ,  \ee
where the old coordinates $X^A$ now depend on new coordinates
\be \tilde X^A = A^A_{\ B} X^B + B^{AB} F_{B}(X) \ . \ee
Finally, we note that a duality transformation also acts as
\be 
\left( \begin{array}{cc} \tilde X^A  &  D_{\bar I} \widetilde{\bar X}^A \end{array} \right) =
\big( A + B {\cal N} \big)^{A}_{\ B} \left( \begin{array}{cc}  X^B  &  D_{\bar I} \bar X^B \end{array} \right) \ . 
\ee

We now  consider the four-dimensional theory specified by the prepotential (\ref{vspecial}) obtained from dimensional 
reduction,  and expand the Lagrangian for the generic Jordan family around the base point 
$c^I$ of the five-dimensional parent theory while introducing the special coordinates $z^I$ as follows,
\be 
Z^A = \Big( 1, {i \over \sqrt{2}} c^I  + z^I  \Big) \ . 
\ee
With this choice, all scalar fields vanish at the base point; the standard choice of $c^I$ for the generic Jordan family is 
given by equation (\ref{base-generic}).  At the base point, we have a canonically normalized scalar metric 
$g_{I \bar J} = \delta_{I  J}$ and a matrix ${\cal N}_{AB}$ given by
\be 
{\cal N}_{AB} = -\text{diag}\Big( {i \over 4}, i, \ldots ,i  \Big) \ . 
\ee
We encounter however difficulties with interpreting the  
supersymmetry transformations (\ref{4dsusy}). Indeed one may see that at the base point \eqref{base-generic} 
the $(\tilde n + 2 )  \times( \tilde n + 2)$ matrix $\big( X^A \ \ D_{\bar I} \bar X^A \big)^t$, which appears in 
the supersymmetry transformations of the fermionic fields (\ref{4dsusy}) is not diagonal and presents some imaginary 
entries. This implies that both field strengths and dual field strengths of spectator fields appear in  the linearized 
supersymmetry variations of the fermionic fields.\footnote{Note that at the base point $G^+_{A} = i F^{A+} $.} 

To make contact between scattering amplitudes computed from the supergravity Lagrangian with the ones 
obtained employing a double-copy construction (which we shall detail in later sections),  it is desirable to go to a symplectic frame 
in which  (1) supersymmetry acts diagonally 
at the base point without mixing fields with different matter indices $I,J=0,1,\ldots,\tilde n$ (so that scalars and vectors with the same 
matter index belong to the same supermultiplet)
and (2) the cubic couplings of the theory are invariant under the maximal compact subgroup $SO(\tilde n)$ of the $U$-duality group of the ungauged theory 
(and hence $SO(\tilde n)$ is a manifest symmetry of the resulting scattering amplitudes).   
It turns out this can be achieved in three steps. 
\begin{enumerate}
\item We first dualize the extra spectator coming from dimensional reduction, $F^{-1}_{\mu \nu}$. 
Using the language introduced at the beginning of this section, we 
 use a duality transformation defined by
\be A = D = \left( \begin{array}{cc} 0 & 0 \\0 & I_{\tilde n +1} \end{array} \right) \ , \qquad 
B = - C = \left( \begin{array}{cc} 1 & 0 \\0 & 0_{\tilde n +1}   \end{array} \right) \ .
 \ee
After this duality transformation, we have the following expressions, 
\be \left( \begin{array}{c} X^A \\ D_{\bar I} \bar X^B \end{array} \right) = J_1 O J_2 \ , \qquad  {\cal N}_{AB} = - i ( J_2)^{-2} \ ,\ee
where $O$ is an orthogonal matrix that acts non-trivially only on the spectator fields and $J_1, J_2$ are diagonal matrices,
\be J_1 = {1\over \sqrt{2}} \text{diag} \big( - i, 1, \ldots ,1 \big) \ , \qquad  J_2 = \text{diag}\Big( {1\over 2}, 1, \ldots ,1 \Big) \ .  \ee

\item
To obtain diagonal supersymmetry transformations at the linearized level, a second $Sp(2{\tilde n}+4, \mathbb{R})$ transformation is necessary: 
\be A = O (J_2)^{-1} \ , \qquad D = O J_2 \ , \qquad  B=C=0 \ . \ee
Note that this transformation, having $B=C=0$, does not involve 
the dualization of any field and can be thought of as a mere field redefinition involving the three vector spectator fields.\footnote{We recall 
that, for simplicity, we restricted ourselves to theories with the minimal number of spectator fields. The discussion here can be generalized without difficulty 
to more spectators.}
After this redefinition the supersymmetry transformations act diagonally with respect to the matter vector indices $I,J=0.1. \ldots, \tilde n$.   
We obtain a simple expression
for the period matrix $\cN_{AB}$,
\bea {\cal N}_{AB} &=& \left( \begin{array}{cc} - i & 2 z^J \\
2 z^I & -i - {4 \over \sqrt{3}} C_{IJK} \bar z^K   \end{array} \right) + \ldots \ \ , 
\eea
where the $C$ tensor is the one corresponding to the generic Jordan family, given by (\ref{Creg}).

\item We finally dualize the extra spectator vector $F^1_{\mu \nu}$ employing a transformation specified by the matrices 
\be A = D = \left( \begin{array}{ccc} I_2 & 0 & 0 \\ 0 & 0 & 0 \\ 0 & 0 & I_{\tilde n -1}  \end{array} \right) \ , \qquad 
B = - C = \left( \begin{array}{ccc} 0_2 & 0 & 0 \\ 0 & 1 & 0 \\ 0 & 0 & 0_{\tilde n -1}  \end{array} \right)   \ .
 \ee
In order to avoid an additional factor of $i$ in the supersymmetry transformation of the scalar field $z^1$, we need to accompany this last duality transformation with
the field redefinition
\be z^1 \rightarrow - i z^1 \ . \label{redefinez} \ee
In the end, the period matrix assumes the following expression up to linear terms in the scalar fields 
\bea {\cal N}_{AB} &=& \left( \begin{array}{cc} - i & 2 z^J \\
2 z^I & -i + {4 \over \sqrt{3}} \tilde C_{IJK} \bar z^K   \end{array} \right) + \ldots \ \ , 
\eea
where we have defined a new tensor $\tilde C_{IJK}$ with non-zero entries 
\be 
\tilde C_{0ab} = {\sqrt{3} \over 2} \delta_{ab} \ , \qquad a,b=1,2,\ldots, \tilde n \ .  
\ee
We note that $\tilde C_{IJK}$ is manifestly invariant under the  $SO(\tilde n)$ symmetry.\footnote{We stress that only $SO(\tilde n-1)$ 
is linearly realized in the symplectic frame we have chosen and that only the cubic vector-vector-scalar couplings posses the extended $SO(\tilde n)$ 
symmetry. To reach a symplectic frame in which the full $SO(\tilde n)$ is linearly realized (such as the one in ref.~\cite{Ceresole:1995jg}), a further nonlinear field redefinition 
is necessary. Since this redefinition becomes the identity map when nonlinearities are removed, it does not affect the $S$ matrix, 
which is already  invariant under $SO(\tilde n)$ transformations. 
}
\end{enumerate}

In appendix \ref{appB} we collect the expansions for the period matrix $\cN_{AB}$, the scalar metric $g_{I \bar J}$ and the K\"{a}hler potential ${\cal K}$ in the 
symplectic frame specified above and  up to quadratic terms 
in the scalar fields.   

The final action for a YMESGT  with compact 
gauge group obtained from the generic Jordan family takes on the following form,
\be e^{-1} {\cal L}= -{1\over 2} R - g_{I \bar J} {\cal D}_\mu z^I {\cal D}^\mu \bar z^J + {1 \over 4} \Im \cN_{AB} {\cal F}^A_{\mu \nu } {\cal F}^{B \mu \nu} 
- {e^{-1} \over 8} 
\epsilon^{\mu\nu\rho\sigma} \Re \cN_{AB} {\cal F}^A_{\mu \nu} {\cal F}^B_{\rho \sigma} + g^2 {\cal P}_4 \ ,  \label{L4dfinal}  \ee
where the gauge covariant derivatives are standard,
\begin{eqnarray}
\mathcal{D}_{\mu} z^{I} & \equiv  & \partial_{\mu} z^{I} + g A_{\mu}^{J}f_{JK}^{I} z^{K} \ , \\
\mathcal{F}_{\mu\nu}^{I}  &  \equiv  &    2\partial_{[\mu}A_{\nu]}^{I} + gf_{JK}^{I}A_{\mu}^{J}A_{\nu}^{K} \ , 
\end{eqnarray}
with $g$ denoting the gauge coupling and $f^{rst}$ the group structure constants.
The four-dimensional potential term ${\cal P}_4$ is given by
\be 
{\cal P}_4 = -  e^{\cal K} g_{rs}f^{rtu}f^{svw} z^t \bar z^u z^v \bar z^w \ .
\ee
As mentioned previously, this is the expected form of the scalar potential, based on the fact that in the limit of
vanishing gravitational constant the YMESGTs reduce to the dimensional reduction of a five-dimensional gauge theory.

It should be noted that the duality transformation that we have employed does not touch fields charged under the gauge group.
By employing the Lagrangian (\ref{L4dfinal}) and the expansions collected in appendix \ref{appB} 
it is straightforward to derive the Feynman rules used to 
obtain the amplitudes presented in the following sections.

\newpage

\section{Color/kinematics duality and the double-copy ${\cal N}=2$ YMESGTs \label{CKsec}}

In section~\ref{general constraints double-copy}  we discussed the properties of gauge theories that can generate, through 
the double-copy construction, minimal couplings between non-abelian gauge fields, spin-$0$ and spin-$1/2$ matter fields. 
In this section we expand that discussion and identify the two gauge theories whose double copy can yield the generic Jordan 
family of YMESGTs in $D=4,5$ dimensions. One of them is the standard ${\cal N}=2$  sYM theory and the other is a particular scalar-vector theory;
we will demonstrate that the amplitudes of the latter obey 
color/kinematics duality through at least six points. Thus, even though we will not construct its higher-point amplitudes in a form manifestly obeying the
duality, they may be used in the double-copy construction. 
Our construction can be carried out in any dimension in which the half-maximal sYM theory exists. In four and 
five dimensions it yields the generic Jordan family of YMESGTs; we shall focus on the four-dimensional case  because of the
advantage provided by the spinor-helicity formalism. The six-dimensional supergravity generated by our construction contains
a graviton multiplet, a $\cN=(1,0)$ self-dual tensor multiplet and ${\tilde n}-2$ Yang-Mills multiplets.  The $6D$ supergravity theories that one obtains by compactifying heterotic string over a $K3$ surface belong to this family of theories coupled to hypermultiplets. Modulo the coupling to hypermultiplets they reduce to the five- 
and four-dimensional generic Jordan family theories (as one can easily see at the level of their scattering amplitudes).\footnote{We 
should however note that generic Jordan family of $5D$ MESGTs can also be obtained from $6D$ , $\cN=(1,0)$ supergravity coupled to  arbitrary number 
$\tilde{n}-1$ of $\cN=(1,0)$ self-dual tensor multiplets. However interacting non-abelian theories of tensor fields are  not known. 
Therefore it is not clear how one can extend  our  results  to such interacting non-abelian tensor theories. }

\subsection{The two gauge-theory factors}

To identify the relevant gauge theories we begin by satisfying the constraints imposed in the 
vanishing gauge coupling limit by the corresponding MESGT, {i.e.} that the  asymptotic spectrum is 
a sum of tensor products of vector and matter multiplets. 
Supergravities of this sort, which may be embedded  in ${\cal N}=8$ supergravity and have at least minimal supersymmetry, as well as general 
algorithms for their construction, have been discussed in refs.~\cite{Damgaard:2012fb, Carrasco:2012ca}. Extensions of these theories to include further 
matter ({i.e.} vector and chiral/hyper multiplets) have also been discussed. 
Moreover, theories whose spectra are truncations of sums of tensor products of matter multiplets have been discussed in refs.~\cite{Chiodaroli:2013upa, Johansson:2014zca}.
It is not difficult to see that the on-shell spectrum of the generic Jordan family of MESGTs may be written as
the tensor product of an ${\cal N}=2$ vector multiplet 
with a vector and $\tilde n$ real scalar fields,
\be
\{A_+, \phi^\allphi{A}, A_-\}\otimes \{A_+, \lambda_+, \varphi, {\bar\varphi},\lambda_-, A_-\} \ ,
\label{2copy_spectrum_ini}
\ee
with real scalars $\phi^\allphi{A}$, $\allphi{A}=1,\dots, \tilde n$. 

Unlike ${\cal N}> 2$  supergravity theories, supergravities with ${\cal N}\leq 2$ are not uniquely specified by their field content.
Since ${\cal N}=2$ MESGTs are specified by their trilinear couplings, to identify the correct double-copy construction, it suffices to 
make sure that the trilinear interaction terms around the standard base point are correctly reproduced. Detailed calculations for 
MESGTs with various numbers of vector multiplets as well as general constructions of such theories as orbifold truncations of 
${\cal N}=8$ supergravity imply that the relevant gauge theories are ${\cal N}=2$ sYM theory and a Yang-Mills-scalar theory that is 
the dimensional reduction of $D=4+{\tilde n}$ pure YM theory.

Starting with such a pair of gauge theories for some number ${\tilde n}$ of scalar fields, 
the next task is to modify one of them such that a $S$-matrix element originating 
from a minimal coupling of supergravity fields is reproduced by the double-copy construction. As discussed in 
section~\ref{LagrangianDiscussion}, from a Lagrangian perspective we may contemplate gauging a subgroup $K$  of the compact 
part of the off-shell global symmetry  group $G$ of the theory. For four-dimensional theories in the generic Jordan family this is 
\be
K\subset G = SO({\tilde n})
\qquad 
{\rm dim}(K) \le {\tilde n} \ .
\ee

The manifest global symmetry of the double-copy construction is the product of the global symmetry groups of two gauge-theory 
factors, $G_L\otimes G_R$. In general, this is only a subgroup of the global symmetry group $G$ of the resulting supergravity theory. 
Since the non-manifest generators act simultaneously on the fields of the two gauge-theory factors,  it is natural to expect that 
such a formulation allows only for a gauge group of the type
\be
K\subset G_L\otimes G_R\subset G \ .
\ee 
Certain supergravity theories admit two (or perhaps several) different double-copy formulations and 
the manifest symmetry group of each of them may be different and each of them may allow for a different gauge group. 
A simple example is ${\cal N}=4$ supergravity coupled to two vector multiplets. If realized as the double-copy 
of two ${\cal N}=2$ sYM theories it exhibits no manifest global symmetries (apart from $R$-symmetry). If realized 
as the product of ${\cal N}=4$
sYM and YM theory coupled to two scalars, it has a global  $U(1)$ symmetry rotating the two scalars into each other, which may 
in principle be gauged.

The double-copy construction of MESGTs in the generic Jordan family described above has a manifest $SO(\tilde n)$ symmetry
rotating the $\tilde n$ scalars into each other. This is  part of the maximal compact subgroup of the Lagrangian 
(albeit not in a prepotential formulation).\footnote{This construction departs slightly from the theories reviewed in section~\ref{Lagrangians4}. In that case it was 
assumed that there are always three spectator fields. While it is possible to make such an assumption here as well, it does not 
simplify any of the considerations; therefore we shall not make it.}

Following the discussion in section~\ref{general constraints double-copy}, to generate the minimal coupling of YMESGTs  between 
scalars, spin-$1/2$ fermions and non-abelian gauge fields it is necessary  that one of the two gauge-theory factors contains a 
dimension-three operator (in $D=4$ counting).  
Since the minimal coupling is proportional to the supergravity gauge-group structure constants, the desired gauge-theory operator  
should be proportional to it as well. If only a subgroup of the manifest symmetry is gauged then only a subset of gauge-theory 
scalars appear in this trilinear coupling; in such a situation the global symmetry of the theory is broken to the subgroup leaving 
the trilinear coupling invariant. The scalars transforming in its complement should lead to the supergravity spectator fields.\footnote{ 
This relation is somewhat reminiscent of the AdS/CFT correspondence, where supergravity gauge fields are dual to 
conserved currents for global symmetries.} 

We are therefore led to the following two Lagrangians (using mostly-minus metric):
\begin{align}
L_{{\cal N}=2}^{D=4} \,=\,& -\frac{1}{4} F_{\mu\nu}^{\hat a}F^{\mu\nu}_{\hat a} 
+(\overline{D_\mu \phi})^{\hat a}(D^\mu\phi)_{\hat a}
-\frac{\SYMcoupling^2}{2} (i\rf_{\hat a \hat b \hat c} \phi^{\hat b}{\bar \phi}^{\hat c})
(i \rf^{\hat a}{}_{\hat b' \hat c'} \phi^{ \hat b'}{\bar \phi}^{\hat c'})
\cr
&\null +i{\bar \lambda}D_\mu{\bar\sigma}^\mu \lambda
+{\SYMcoupling \over\sqrt{2}}  (i\rf_{\hat a \hat b \hat c })(\lambda^{\hat a\alpha \xi}{\phi}^{\hat b}\lambda^{\hat c \beta}_\xi)\epsilon_{\alpha\beta}
+{\SYMcoupling \over\sqrt{2}} (i\rf_{\hat a \hat b \hat c })({\bar \lambda}^{\hat a \alpha {\dot \xi}}{\bar \phi}^{\hat b}{\bar\lambda}^{\hat c\beta}_{\dot \xi})\epsilon_{\alpha\beta} \,,
\end{align}
where $\alpha, \beta \ $ are $SU(2)$ indices and
\bea
L_{{\cal N}=0} &=& -\frac{1}{4} F_{\mu\nu}^{\hat a}F^{\mu\nu}_{\hat a} 
+ \frac{1}{2}(D_\mu\phi^{\allphi{A}})^{\hat a}(D^\mu\phi^{\allphi{ B}})_{\hat a}\delta_{\allphi{A}\allphi{B}}
+\frac{\SYMcoupling^2}{4} (i\rf_{\hat a \hat b \hat c } \phi^{\hat b\allphi{B}}\phi^{\hat c \allphi{C}})(i\rf^{\hat a}{}_{\hat b' \hat c'} \phi^{\hat b'\allphi{B}'}
\phi^{\hat c'\allphi{C}'})\delta_{\allphi{B} \allphi{B}'}\delta_{\allphi{C}\allphi{C}'}\cr
&&   +\frac{\SYMcoupling g'}{3!} (i\rf_{\hat a \hat b \hat c }) F_{\allphi{A}\allphi{B}\allphi{C}} \phi^{\hat a\allphi{A}} \phi^{\hat b\allphi{B}} \phi^{\hat c\allphi{C}}
\ , \label{vectorscalarL}
\\
F_{\mu\nu}^{\hat a}&=& \partial_\mu A_\nu^{\hat a} - \partial_\nu A_\mu^{\hat a} 
+ \SYMcoupling \ \rf^{\hat a}{}_{\hat b \hat c}A^{\hat b}_\mu A^{\hat c}_\nu \ ,
\label{YMStheory}
\qquad
(D_\mu\phi^{a})^{\hat a} = \partial_\mu \phi^{ {\hat a a}}  
+ \SYMcoupling \ \rf^{\hat a}{}_{\hat b \hat c}A^{\hat b}_\mu \phi^{\hat c{ a}} \ ,
\eea
where the gauge-group generators are assumed to be hermitian, $[T_{\hat a},T_{\hat b}]= i\rf_{{\hat a}{\hat b}}{}^{\hat c}T_{\hat c}$ and 
the coefficient $g'$ is arbitrary and dimensionful.\footnote{The gauge-group structure constants $\rf$ are related to the structure 
constants ${\tilde \rf}$, naturally appearing in color-dressed scattering amplitudes, through
$$
\rf_{\hat a \hat b \hat c } =-\frac{i}{\sqrt{2}} {\tilde \rf}_{{\hat a} {\hat b} {\hat c}} \ .
$$
} 
The indices $\allphi{A}, \allphi{B}. \ldots$  take values $1,\dots, \tilde n.$
The rank-three tensor $F$ has  entries $F^{rst}$, with $r,s,t = 2,3,\ldots, \tilde n$, given 
by the structure constants 
of a subgroup $K$ of $SO(\tilde n)$, and all other entries set to zero.\footnote{A possible proportionality coefficient is absorbed in the coefficient $g'$.}
To use the double-copy construction we need the scattering amplitudes of both the ${\cal N}=2$ sYM theory  and the 
Yang-Mills-scalar theory \eqref{vectorscalarL} to obey color/kinematics duality, albeit only one of them is needed in a form that
obeys it manifestly. Since the amplitudes of the former theory have this property (and their manifestly color/kinematic-satisfying
form may be obtained by a $\mathbb{Z}_2$ projection from the corresponding amplitudes of ${\cal N}=4$ sYM theory) we only need 
to make sure that the vector/scalar theory obeys the duality as well. We shall explore this question in the next subsection with a 
positive conclusion. 

Denoting by 
\be
{\rm G}_+ = g_+ +\eta_\alpha \lambda_+^\alpha+\eta^2 \phi
~,\qquad
{\rm G}_- = {\bar \phi}+\eta_\alpha \lambda_-^\alpha+\eta^2g_-
 \ee
the two $CPT$-conjugate on-shell vector multiplets of ${\cal N}=2$ sYM, the on-shell ${\cal N}=2$ multiplets of the supergravity
theory are
\be
\begin{array}{lcl}
{\rm H}_+ =A_+\otimes {\rm G}_+ = h_{++}+\eta_\alpha \psi_+^\alpha +\eta^2 V_+ \ ,
&&
{\rm H}_- = A_-\otimes {\rm G}_- = V_- +\eta_\alpha \psi^\alpha_- +\eta^2 h_{--}  \ ,
\\
{\widetilde {\rm V}}_+ = A_+\otimes {\rm G}_- =  {\tilde V}_+   +\eta_\alpha {\tilde \zeta}_+^{\alpha} +\eta^2  S_{+-} \ ,
&&
{\widetilde {\rm V}}_- = A_-\otimes {\rm G}_+ =  S_{-+} +\eta_\alpha {\tilde \zeta}_-^{\alpha} + \eta^2 {\tilde V}_- \ ,
\\
{\rm V}_+^{\specphi{M}} = \phi^{\specphi{M}} \otimes {\rm G}_+ = V_+^{\specphi{M}}   +\eta_\alpha \zeta_+^{\specphi{M}\alpha} +\eta^2 _+ S^{\specphi{M}} \ ,
&&
{\rm V}_-^{\specphi{M}} = \phi^{\specphi{M}} \otimes {\rm G}_- ={\bar S}^{\specphi{M}}   +\eta_\alpha \zeta_-^{\specphi{M}\alpha} +\eta^2  V_-^{\specphi{M}} \ ,
\\
{\rm V}_+^r =  \phi^{r} \otimes {\rm G}_+ = V_+^r  +\eta_\alpha \zeta_+^{r\alpha} +\eta^2 _+ S^r \ ,
&&
{\rm V}_-^r =  \phi^{r} \otimes {\rm G}_- ={\bar S}^r   +\eta_\alpha \zeta_-^{r\alpha} +\eta^2  V_-^r \ ,
\end{array} \quad \label{multiplets}\ee
with the index $r$ transforming in the adjoint representation of $K$. 
The component fields are the same as in the corresponding MESGT; denoting 
by subscript $1$ and $2$  the component fields of the ${\cal N}=0$ and ${\cal N}=2$ gauge theory respectively,  they are
\be
\begin{array}{rlcl}
\text{spin}=2: &  h_{++}=A_{1+} \otimes A_{2+} \ , &  ~~~~& h_{--}=A_{1-} \otimes A_{2-} \ ,  \\
 \text{spin}=3/2: &  ~\,
 \psi^\alpha_{+}=A_{1+} \otimes \lambda^\alpha_{2+} \ , &  & ~\,\psi^\alpha_{-}=A_{1-} \otimes \lambda^\alpha_{2-} \ , \\
\text{spin}=1: & ~\,
V_+ =  A_{1+} \otimes \varphi_{2} \ , &&  ~\,{\tilde V}_- =  A_{1-} \otimes \varphi_{2} \ , \\
                      & ~\,
{\tilde V}_+ =  A_{1+} \otimes {\bar \varphi}_{2} \ , && ~\,V_- =  A_{1-} \otimes {\bar \varphi}_{2} \ , \\
                      & ~\,
 V^{\allphi{A}}_+ =  \phi_1^\allphi{A} \otimes A_{2+} \ , && ~V^\allphi{A}_- =  \phi_1^\allphi{A} \otimes A_{2-} \ , \\
\text{spin}=1/2:& ~~{\tilde \zeta}^\alpha_+=A_{1+}\otimes  \lambda^\alpha_{2-} \ , && 
~\,{\tilde \zeta}^\alpha_-=A_{1-}\otimes  \lambda^\alpha_{2+} \ , \\                                      
                        & ~\zeta^{\allphi{A}\alpha}_+=\phi_1^{\allphi{A}}\otimes  \lambda^\alpha_{2+} \ , && 
                        \zeta_-^{\allphi{A}\alpha}=\phi_{1}^{\allphi{A}}\otimes  \lambda^\alpha_{2-} \ , \\               
\text{spin}=0: &  ~\;S^{\allphi{A}} =  \phi_1^{\allphi{A}}\otimes \varphi_2  \ , && 
~{\bar S}^{\allphi{A}} =  \phi_1^{\allphi{A}}\otimes {\bar \varphi}_2  \ , \\
                      & S_{+-} = A_{1+}\otimes A_{2-}  \ ,&& \!\! S_{-+} = A_{1-}\otimes A_{2+} \ .
\end{array}
\nonumber
\label{fields}
\ee
As we shall see in the next section, multiplets carrying indices $r, s, ...$ will be identified with the gauge field multiplets of 
supergravity, while those carrying indices $\specphi{M}, \specphi{N}, \dots$ will be related to the 
supergravity spectator multiplets with the same indices while $V$ and 
${\tilde V}$ will be related to the ``universal" spectator vector fields of the generic Jordan family YMESGTs. 
The resulting theory will have an $SO(\tilde n - \text{dim}(K) )$ global symmetry acting on 
the indices $\specphi{M}, \specphi{N}, \dots$.\footnote{We note that, by simply replacing the ${\cal N}=2$ sYM theory with ${\cal N}=4$ gauge theory, the double-copy theory becomes 
${\cal N}=4$ supergravity coupled with $\tilde n - \text{dim} (K)$ abelian and ${\rm dim}(K)$ non-abelian vector multiplets, which is 
uniquely specified by its field content and symmetries (up to symmetry and duality transformations of abelian vector fields).}
The fields labeling the amplitudes obtained through the double-copy construction need not {\it a priori} be the same as the natural 
asymptotic states around a Minkowski vacuum following from the Lagrangian (\ref{L4dfinal}) and a field redefinition may be required. 
Such redefinitions are to  be constructed on a case-by-case basis, for the specific choice of Lagrangian asymptotic fields. 
As we shall see in section~\ref{explicit}, for the choice of symplectic section in section~\ref{Lagrangians4} the map between 
the Lagrangian and the double-copy asymptotic states is trivial; additional nonlinear field redefinitions such as those 
needed to restore $SO({\tilde n})$ symmetry of the Lagrangian should not affect the $S$ matrix.

\subsection{Color/kinematics duality of Yang-Mills-scalar theories \label{CKvs}}

To use the scattering amplitudes from the Lagrangian \eqref{vectorscalarL} to find supergravity amplitudes either 
through the KLT or 
through the double-copy construction, it is necessary to check that, in principle, they can be put in a form obeying color/kinematics 
duality in $D$ dimensions.\footnote{This is related to the need of regularization at loop level.}
Since at $g'=0$,  the equation~\eqref{vectorscalarL} reduces to the dimensional reduction of a $(4+{\tilde n})$-dimensional pure YM 
theory, which is known to obey the duality; we need to check only $g'$-dependent terms.

For the four-point amplitudes this can be done simply by inspection. The only $g'$-dependent amplitude involving four 
scalars is given by ref.~\cite{Chiodaroli:2013upa}
 \begin{equation}
\begin{aligned}
&\mathcal{A}^{(0)}_4(1^{\phi^{{\allphi{ A}}_1}}2^{\phi^{\allphi{ A}_2}}3^{\phi^{\allphi{ A}_3}}4^{\phi^{\allphi{ A}_4} })\big|_{g'{}^2\text{ terms}}
=\SYMcoupling^2 g'{}^2\left(\frac{1}{s}F^{\allphi{ A}_1\allphi{ A}_2\allphi{ B}}F^{\allphi{ A}_3\allphi{ A}_4\allphi{ B}}  
\rf^{\hat a_1\hat a_2 \hat b}\rf^{\hat a_3 {\hat a}_4 \hat b}
\right.\cr
&\left.\qquad\qquad\qquad
                        +\frac{1}{u}F^{\allphi{ A}_3\allphi{ A}_1\allphi{ B}}F^{\allphi{ A}_2\allphi{ A}_4\allphi{ B}}  
                        \rf^{\hat a_3 \hat a_1\hat b}\rf^{\hat a_2 {\hat a}_4 \hat b}
                        +\frac{1}{t}F^{\allphi{ A}_2\allphi{ A}_3\allphi{ B}}F^{\allphi{ A}_1\allphi{ A}_4\allphi{ B}}  
                        \rf^{\hat a_2 \hat a_3 \hat b}\rf^{\hat a_1{\hat a}_4 \hat b} \right)\ .\\
\end{aligned}
\label{notinpairs}
\end{equation}
We therefore see that color/kinematics duality is satisfied if the nonzero part of $F_{{a}{b}{c}}$ obeys the Jacobi identity and, therefore, 
it is proportional to the structure constants of some group. This 
is consistent with the expectation that the trilinear scalar coupling is responsible for generating the minimal couplings of the supergravity
gauge fields.
One may similarly check that the $g'$-dependent terms in the four-point scalar amplitude 
with pairwise identical scalars have a similar property \cite{Chiodaroli:2013upa}.

An amplitude that probes  the  scalar interactions both on their own as well as together with the scalar-vector interactions 
is $\mathcal{A}^\text{tree}_5(1^{\phi^{{a}_1}}2^{\phi^{{a}_2}}3^{\phi^{{a}_3}}4^{\phi^{{a}_3}}5^{\phi^{{a}_3}})$ 
involving only three distinct scalars. The ${\cal O}(g'{}^3)$ part of this amplitude is
\bea
\label{5Sgp3}
&&\hskip -25pt \mathcal{A}^{(0)}_5(1^{\phi^{{a}_1}}2^{\phi^{{a}_2}}3^{\phi^{{a}_3}}4^{\phi^{{a}_3}}5^{\phi^{{a}_3}})
\big|_{g'{}^3\text{ terms}}
\no \\
&&\qquad
= \SYMcoupling^3g'{}^3 
F^{{a}_1{a}_3{b}} F^{{b}{a}_3{c}} F^{{c}{a}_3{a}_2}
\sum_{\sigma\in {\cal S}(3,4,5) }\frac{1}{s_{1\sigma(3)}s_{2\sigma(5)}} \rf^{\hat a_1\hat a_{\sigma(3)} \hat b}\rf^{\hat b \hat a_{\sigma(4)} \hat c}\rf^{\hat c \hat a_{\sigma(5)} \hat a_2} \ .
\eea
By construction this amplitude obeys color/kinematics duality, or rather ``color/color duality''. Moreover, it is easy to check that it satisfies the BCJ amplitude relations~\cite{Du:2011js}.

The $\mathcal{A}^{(0)}_5(1^{\phi^{{a}_1}}2^{\phi^{{a}_2}}3^{\phi^{{a}_3}}4^{\phi^{{a}_3}}5^{\phi^{{a}_3}})$ receives contributions
from both the cubic and quartic terms of the Lagrangian. 
Defining $k_{ij...}=k_i+k_j+...$ and $k_{i{\bar j}...}=k_i-k_j+...$ it  is given by
\bea
\label{5Sgp1}
&& \hskip -25pt \mathcal{A}^{(0)}_5(1^{\phi^{{a}_1}}2^{\phi^{{a}_2}}3^{\phi^{{a}_3}}4^{\phi^{{a}_3}}5^{\phi^{{a}_3}})
\big|_{g'\text{ terms}} \no
\\
&&=\frac{1}{2}\SYMcoupling^3 g' F^{{a}_1{a}_2{a}_3} \left[\left(
\frac{k_{12{\bar 3}}\cdot k_{4{\bar 5}}}{s_{12}s_{45}} \rf^{\hat a_1\hat a_2 \hat b}\rf^{\hat b \hat a_3 \hat c} 
+ \frac{k_{{\bar 1}23}\cdot k_{4{\bar 5}}}{s_{23}s_{45}} \rf^{\hat a_2 \hat a_3 \hat b}\rf^{\hat b \hat a_1\hat c} 
\right.\right. \cr
&& \qquad\qquad\qquad\qquad\qquad
+  \left.\left. 
\frac{k_{1{\bar 2}3}\cdot k_{4{\bar 5}}}{s_{13}s_{45}}\rf^{\hat a_3 \hat a_1\hat b}\rf^{\hat b \hat a_2 \hat c}\right)\rf^{\hat c {\hat a}_4 {\hat a}_5}
+(3\leftrightarrow 4)+(3\leftrightarrow 5) \right]
\cr
&&+\frac{1}{2}\SYMcoupling^3 g' F^{{a}_1{a}_2{a}_3} \left[
-\left(\frac{1}{s_{13}}+\frac{1}{s_{24}}\right) \rf^{\hat a_1 \hat a_3 \hat b}  \rf^{\hat b {\hat a}_5 \hat c} \rf^{\hat c \hat a_2 {\hat a}_4}
-\left(\frac{1}{s_{13}}+\frac{1}{s_{25}} \right) \rf^{\hat a_1\hat a_3 \hat b}  \rf^{\hat b {\hat a}_4 \hat c} \rf^{\hat c \hat a_2 {\hat a}_5}
\right.\cr
&& \qquad\qquad\qquad\qquad\qquad
\left. \vphantom{\frac{1}{s_{13}}}+(3\leftrightarrow 4)+(3\leftrightarrow 5) \right] \ .
\eea
It is not difficult to check that the color-ordered amplitudes following form this expression obey the five-point amplitudes relation
\eqref{amplituderelations} and its images under permutations of external lines;
it therefore follows that 
there exists a sequence of generalized gauge transformations~\cite{BCJ} that casts  this amplitude into a form manifestly 
obeying color/kinematics duality. Since  ${\cal N}=2$ amplitudes manifestly obeying the duality are known \cite{Chiodaroli:2013upa}, 
for the purpose of constructing YMESGT amplitudes,  it is not necessary to also have a manifest representation for the amplitudes 
of Yang-Mills-scalar theory (albeit it might lead to more structured expressions if one were available).

We have also checked that the tree-level six-point amplitudes following from the Lagrangian \eqref{vectorscalarL} obey the 
relevant amplitude relations \cite{BCJ} and therefore they should also have a presentation manifestly obeying the duality. Beyond six points, we conjecture that the tree amplitudes of \eqref{vectorscalarL} always satisfy the BCJ relations \cite{BCJ}, and thus the theory should satisfy color-kinematics duality at tree level. From this one can expect that it may also satisfy the duality at loop level~\cite{BCJLoop}.

\section{Tree-level amplitudes  \label{explicit}}

Having established that the scattering amplitudes of the Yang-Mills-scalar theory \eqref{vectorscalarL} obey color/kinematics duality, 
we proceed to use it to evaluate explicitly the double-copy three- and four-point amplitudes  and compare them with the analogous 
amplitudes computed from the Lagrangian (\ref{L4dfinal}).

The color structure of supergravity amplitudes is the same as that of a gauge theory coupled to fields that are singlets under 
gauge transformations. In a structure-constant basis they are given by open (at tree-level) and closed (at loop-level) strings 
of structure constants and color-space Kronecker symbols. In the trace basis, this implies that the structure of tree amplitudes
is similar to that of loop amplitudes in that, unlike pure gauge theories, it is not restricted to having only single-trace terms:
\bea
{\cal A}_n&=&\sum_{{\cal S}_m} \; \Tr[T^{a_1}\dots T^{a_m}] A_{n, 1}(1\dots n)
\\
&+&\sum_{m_1+m_2=m}
\sum_{{\cal S}_{m_1}\times {\cal S}_{m_2}} \; \Tr[T^{a_1}\dots T^{a_{m_1}}] \Tr[T^{a_{m_1+1}}\dots T^{a_{m_1+m_2}}] 
A_{n, 2}(1\dots n)+\dots \ ,
\nonumber
\eea
where ${\cal S}_{m_i}$ is the set of non-cyclic permutations. Different traces are ``connected" by exchange of color singlets.

In theories with less-than-maximal supersymmetry, scattering superamplitudes are organized following the 
number of on-shell multiplets the asymptotic states belong to.  In our case, using the fact that ${\cal N}=2$  algebra 
is a $\mathbb{Z}_2$ orbifold of ${\cal N}=4$ algebra we can use a slightly more compact organization.
To this end we organize the supergravity multiplets \eqref{multiplets} as
\bea
{\cal H}_+ &=& {\rm H}_+ + \eta^3\eta^4 {\widetilde{\rm V}}_+  = h_{++}+\eta_\alpha \psi_+^\alpha +\eta^1 \eta^2 V_+ +  
\eta^3 \eta^4 {\tilde V}_+   + \eta^3 \eta^4 \eta_\alpha {\tilde \zeta}_+^{\alpha} + \eta^1 \eta^2 \eta^3 \eta^4  S_{+-} \ , \no  \\
{\cal H}_- &=& {\widetilde {\rm V}}_-  + \eta^3\eta^4{\rm H}_- =   S_{-+} +  \eta_\alpha {\tilde \zeta}_-^{\alpha} + \eta^1\eta^2 {\tilde V}_- + 
\eta^3 \eta^4  V_- + \eta^3 \eta^4  \eta_\alpha \psi^\alpha_- +\eta^1 \eta^2 \eta^3 \eta^4  h_{--}  \ , \no
\\
{\cal V}^{\specphi{ A}} &=& {\rm V}_+^{\specphi{A}} +\eta^3\eta^4{\rm V}_-^{\specphi{A}}  = 
V_+^{\specphi{ A}}  +\eta_\alpha \zeta_+^{\specphi{ A} \alpha} + \eta^1 \eta^2  S^{\specphi{ A}} + \eta^3 \eta^4 {\bar S}^{\specphi{ A}}  
+\eta^3 \eta^4 \eta_\alpha \zeta_-^{\specphi{ A} \alpha} +\eta^1 \eta^2 \eta^3 \eta^4 V_-^{\specphi{ A}} \ ,
\label{constrainedmultiplets} \qquad \eea
and the $\cN=2$ gauge multiplet as
\bea
{\cal G} &=& {\rm G}_+ +\eta^3\eta^4 {\rm G}_- = g_+ + \eta_\alpha \lambda^\alpha_+ + \eta^1 \eta^2 \phi + \eta^3 \eta^4 \bar \phi + \eta^3 \eta^4 \eta_\alpha 
\lambda^\alpha_- + \eta^1 \eta^2 \eta^3 \eta^4 g_- \ , 
\eea
where $\eta^{3,4}$ are auxiliary Grassmann variables.\footnote{Since they always appear as a product one may also replace 
$\eta^3\eta^4$ by a nilpotent Grassmann-even variable.} One may think of ${\cal H}_\pm$ and ${\cal V}$ as constrained 
${\cal N}=4$ supergravity and vector multiplets, respectively, which are invariant under the $\mathbb{Z}_2$ projection.\footnote{One may 
find the amplitudes of ${\cal N}=4$ supergravity coupled to $n_s$ abelian and ${\rm dim}(K)$ non-abelian vector multiplets by simply 
forgetting the $\mathbb{Z}_2$ projection.}
With the asymptotic states assembled in these superfields,  superamplitudes are polynomials in the pairs $\eta^{3}_i\eta^{4}_i$ 
with $i$ labeling the external legs. The monomial with $n_+$ such pairs represents the superamplitudes with $n_+$ supermultiplets 
of type $+$.

\subsection{Three-point amplitudes and the field and parameter map \label{3pf_field_parameter_map}}

Three-point amplitudes verify  the structure of minimal couplings and of other trilinear couplings demanded supersymmetry 
and consistency of the YMESGT, such as the reduction to four dimensions of the fermion bilinear \eqref{gaugerelatedfermionbilinear}.
They also determine the map between the double-copy and Lagrangian fields and parameters.

The kinematic parts of the ${\cal N}=0$ amplitudes involving  at least one gluon are the same as in ${\cal N}=4$ sYM; the three-scalar 
amplitude -- the only three-point amplitude dependent on $g'$ -- is momentum-independent. Up to conjugation and relabeling of 
external legs, the non-vanishing amplitudes of the Yang-Mills-scalar theory are 
\bea
\label{ggg}
{\cal A}^{(0),{\cal N}=0}_3(1g_-^{\hat a},2g_-^{\hat b},3g_+^{\hat c})&=&
i \SYMcoupling \frac{\langle 12\rangle^3}{\langle 23\rangle\langle 31\rangle}{\tilde \rf}^{\hat a \hat b \hat c} \ ,
\\
\label{SSS}
{\cal A}^{(0),{\cal N}=0}_3(1\phi^{a \allphi{A}},2\phi^{\hat b \allphi{B}},3\phi^{\hat c\allphi{C}})&=&
\frac{i}{\sqrt{2}} \SYMcoupling \, g' {\tilde \rf}^{\hat a \hat b \hat c}{F}^{abc} \ ,
\\
\label{SSg}
{\cal A}^{(0),{\cal N}=0}_3(1\phi^{\hat a \allphi{A}},2\phi^{\hat b{\allphi{B}}},3g_-^{\hat c})&=&
 i\SYMcoupling \frac{\langle 23\rangle \langle 31 \rangle}{\langle 12 \rangle}{\tilde \rf}^{\hat a \hat b \hat c}\delta^{{a}{b}} \ ,
\eea
where ${\tilde \rf}^{\hat a \hat b \hat c} = i\sqrt{2} \rf^{\hat a \hat b \hat c }$.

The ${\cal N}=2$ superamplitudes, labeled in terms of the multiplets ${\cal G}$, may be obtained from those of ${\cal N}=4$ sYM 
theory through the supersymmetric $\mathbb{Z}_2$ orbifold projection acting on the $\eta^3$ and $\eta^4$ Grassmann variables. 
This effectively amounts to modifying the super-momentum conservation constraint as
\be
\label{projection}
\delta^{(8)}(\sum \eta_i^\alpha |i\rangle)\mapsto -{\cal Q}^{34}_n\;\delta^{(4)}(\sum \eta_i^\alpha |i\rangle) \ ,
\qquad
\delta^{(4)}(\frac{1}{2}\sum_{i,j,k}\epsilon_{ijk}[i j] \eta_k^\alpha) \mapsto {\widetilde{\cal Q}^{34}}_3\;\delta^{(2)}(\frac{1}{2}\sum_{i,j,k}\epsilon_{ijk}[i j] \eta_k^\alpha) 
\ee
where 
\be
{\cal Q}^{34}_n = \sum_{1\le i < j\le n}\langle ij\rangle^2(\eta_i^3\eta_i^4)(\eta_j^3\eta_j^4) 
~~~~~~\text{and}~~~~~~
{\widetilde{\cal Q}}^{34}_3=\frac{1}{2}\sum_{i\ne j\ne k=1}^3[ ij]^2(\eta_k^3\eta_k^4) \ .
\ee
Of course, for higher-multiplicity amplitudes other projected supersymmetry invariants appear as well.
With this notation, the three-point superamplitudes are
\be
\begin{aligned}
{\cal A}^{(0),\text{MHV}, {\cal N}=2}_3(1^{\cal G},2^{\cal G},3^{\cal G}) &= - i \SYMcoupling 
{\tilde \rf}^{\hat a \hat b \hat c} \; {\cal Q}^{34}_3\;
\frac{\delta^{(4)}(\sum \eta_i^\alpha |i\rangle)}{\langle 12\rangle\langle 23\rangle\langle 31\rangle}
\ , \cr
{\cal A}^{(0),\overline{\text{MHV}}, {\cal N}=2}_3(1^{\cal G},2^{\cal G},3^{\cal G}) &= i\SYMcoupling 
{\tilde \rf}^{\hat a \hat b \hat c}\; {\widetilde{\cal Q}}^{34}_3\;
\frac{\delta^{(2)}(\frac{1}{2}\sum_{i,j,k}\epsilon_{ijk}[i j] \eta_k^\alpha)}{[12][23][31]} \ .
\end{aligned}
\label{3ptNeq2}
\ee
The two superamplitudes \eqref{3ptNeq2} are related by conjugation and Grassmann-Fourier transform.
From the perspective of the ${\cal N}=4$ theory, ${\cal Q}^{34}_n$ and ${\widetilde{\cal Q}}^{34}_3$ are 
the $\mathbb{Z}_2$-invariant combination of the $\eta^3$ and $\eta^4$ Grassmann variables.\footnote{\label{extractsupercomponents}
To extract  from equation~\eqref{3ptNeq2} 
scattering amplitudes labeled by the ${\cal N}=2$ multiplets ${\rm G}_\pm$ one simply extracts the coefficients of the various 
monomials in $\eta^3\eta^4$. For example,
\bea
{\cal A}^{(0),\text{MHV}, {\cal N}=2}_3(1^{{\rm G}_+},2^{{\rm G}_-},3^{{\rm G}_-}) &=& 
-i\SYMcoupling {\tilde \rf}^{\hat a \hat b \hat c} \;
\frac{ \langle 23\rangle^2}{\langle 12\rangle\langle 23\rangle\langle 31\rangle}\delta^{(4)}(\sum \eta_i^\alpha |i\rangle) \ ,
\cr
{\cal A}^{(0),{\overline{\text{MHV}}}, {\cal N}=2}_3(1^{{\rm G}_+},2^{{\rm G}_+},3^{{\rm G}_-}) &=&
i\SYMcoupling {\tilde \rf}^{\hat a \hat b \hat c}\;
\frac{[12]^2}{[12][23][31]} \delta^{(2)}(\frac{1}{2}\sum_{i,j,k}\epsilon_{ijk}[i j] \eta_k^\alpha) \ .
\nonumber
\eea
}

Using  equations~\eqref{3ptNeq2} and \eqref{SSS}, \eqref{SSg} and \eqref{ggg} it is easy to construct the double-copy  
three-point amplitudes; some of them vanish identically because of special properties of three-particle complex momentum 
kinematics (which {e.g.} implies that the product of holomorphic and anti-holomorphic spinor products vanishes identically).
Up to conjugation, the non-vanishing superamplitudes are 
\be
\begin{aligned}
{\cal M}_3^{(0)}(1^{{\cal H}_-},2^{{\cal H}_-},3^{{\cal H}_+})&= i\frac{\kappa}{2} 
\frac{\langle 12\rangle^2}{\langle 23\rangle^2\langle 31\rangle^2}\;
{\cal Q}^{34}_3\;\delta^{(4)}(\sum \eta_i^\alpha |i\rangle) \ ,
\\
{\cal M}_3^{(0)}(1^{{\cal V}^\allphi{A}},2^{{\cal V}^\allphi{B}},3^{{\cal V}^\allphi{C}})&= i\frac{\kappa}{2\sqrt{2}} 
\frac{g' {F}^{\allphi{A}\allphi{B}\allphi{C}}}{\langle 12\rangle\langle 23\rangle\langle 31\rangle}\;
{\cal Q}^{34}_3\;\delta^{(4)}(\sum \eta_i^\alpha |i\rangle) \ ,
\\
{\cal M}_3^{(0)}(1^{{\cal V}^\allphi{A}},2^{{\cal V}^\allphi{B}},3^{{\cal H}_-})&= i \frac{\kappa}{2} 
\frac{\delta^{\allphi{A}\allphi{B}}}{\langle 12\rangle^2}\;
{\cal Q}^{34}_3\;\delta^{(4)}(\sum \eta_i^\alpha |i\rangle) \ .
\end{aligned}
\label{the3ptAmplitudes}
\ee
The superamplitudes labeled by ${\cal N}=2$ on-shell supermultiplets may be extracted as shown in 
footnote~\ref{extractsupercomponents}.
The component amplitudes extracted from these superamplitudes and their conjugates are very similar to
the component amplitudes following from the supergravity Lagrangian~\eqref{L4dfinal}.\footnote{
To compare the amplitudes from the Lagrangian with those from the double-copy, 
we also need to employ analytic continuation, as the former are obtained with a mostly-plus 
metric and the latter with a mostly-minus metric.}
Indeed, the kinematic factors are fixed by little-group scaling and gauge invariance and the numerical coefficients
can be mapped into each other by identifying the structure constants of the global symmetry group of the 
Yang-Mills-scalar  theory with the structure constants of the supergravity gauge group as\footnote{We may formally 
separate the identification of the gauge coupling from that of the structure constants through the relation
\be
\SYMcoupling^{2m} (g')^l \mapsto \left(\frac{\kappa}{2}\right)^{2m}g^l \ .\no
\ee
}
\be
g'{F}^{rst} = 2ig
{f}^{rst} \ . \label{Fmap}
\ee
All the double-copy and Lagrangian multiplets are then trivially mapped into each other,\footnote{
More in general, one can introduce a parameter $\theta$ in the identification of the multiplets,
\be
\begin{array}{lclcl}
h_{\pm\pm} =h_{\pm\pm}^{\cal L}
&&
V_\pm = \pm i e^{\mp i\theta} V_\pm^{\cal L}
&&
\\
{\tilde V}_\pm = \pm i e^{\pm i\theta}{\tilde V}_\pm^{\cal L}
&&
S_{+-} = - i S_{+-}^{\cal L}
&&
S_{-+} = + i S_{-+}^{\cal L}
\\
{V}_\pm^{\specphi{A}}  =  { V}^{\specphi{A}} {}_\pm^{\cal L}
&&
S^{\specphi{A}}  =   e^{-i\theta} S^{\specphi{A}} {}^{ \cal L}
&&
{\bar S}^{\specphi{A}} =  e^{i\theta} {\bar S}^{\specphi{A}} {}^{ \cal L} 
\\
{V}_\pm^{r} = { V}^{ r}{}_\pm^{\cal L}
&&
S^{r} =  e^{-i\theta}S^{r}{}^{ \cal L}
&&
{\bar S}^{r} =   e^{i\theta}{\bar S}^{r}{}^{ \cal L} 
\end{array} \no
\ee
The parameter $\theta$ is free and its presence is a reflection of the classical $U(1)$ electric/magnetic duality of the theory.
The choice 
$\theta=\pi/2$  is a consequence 
of the symplectic section chosen in section~\ref{Lagrangians4}.} 
\be
\begin{array}{lcl}
{\rm H}_+ = {\rm H}_+^{{\cal L}} \ ,
&~~~~~&
{\rm H}_- = {\rm H}_-^{{\cal L}} \ ,
\\
{\widetilde {\rm V}}_+ = {\widetilde {\rm V}}_+^{\cal L} \ ,
&&
{\widetilde {\rm V}}_- =  {\widetilde {\rm V}}_-^{\cal L} \ ,
\end{array}
~~~~~~~
\begin{array}{lcl}
{\rm V}_+^{\specphi{A}} =  {\rm V}^{\cal L}{}_+^{\specphi{A}} \ , 
&~~~~~&
{\rm V}_-^{\specphi{A}} = {\rm  V}^{\cal L}{}_-^{\specphi{A}} \ ,
\\
{\rm V}_+^r = {\rm V}^{\cal L}{}_+^r \ ,
&&
{\rm V}_-^r = {\rm V}^{\cal L}{}_-^r \ ,
\end{array} 
\label{multiplet_map}
\ee
where the field map is presented in terms of on-shell superfields and  we have added 
the superscript ''${\cal L}$" to the superfields from the supergravity Lagrangian (\ref{L4dfinal}).
The above identity map also shows that the 
linearized Lagrangian and double-copy supersymmetry generators are essentially the same.
The Lagrangian matter multiplets are\footnote{The factor of $i$ can be understood by noticing that in ${\cal N} = 8$ supergravity dilatonic scalars are of the form 
$\phi^{ABCD} + \phi_{ABCD}$ while in ${\cal N} = 2$ language the dilatonic scalars are the imaginary parts of the 
complex scalars that appear in the Lagrangian.}  
\be
{\rm V}^{\cal L}_+ = V^{\cal L}_+ + \eta_\alpha \zeta_+^{\cal L}{}^\alpha +i\eta^2 S^{\cal L} \ .
\ee
We note here that the second amplitude \eqref{the3ptAmplitudes} and its $CPT$-conjugate contain the $S$-matrix element originating 
from the four-dimensional analog of the Yukawa interaction \eqref{gaugerelatedfermionbilinear}, which is required by consistency 
of gauge interactions and supersymmetry. The fact that this amplitude correctly reproduces the interaction is a strong indication that the double-copy 
construction proposed here is capturing correctly all features of the generic Jordan family of four-dimensional YMESGTs.

We also note that the map above establishes an off-shell double-copy structure for the minimal couplings of YMESGTs. Indeed, 
the off-shell double-copy of the three-scalar vertices of the Yang-Mills-scalar theory and the vertices of the ${\cal N}=2$ sYM theory 
simply replaced the color structure constants of the latter with the structure constants of the YMESGT gauge group.
This is consistent the YMESGTs minimal couplings of a standard  ${\cal N}=2$ gauge theory, as it can be seen
trivially in the $\kappa\rightarrow 0$ limit.

\subsection{Four-point amplitudes \label{fourpttree}}

To reinforce the validity of the construction for 
YMESGTs we  proceed to compare the four-point amplitudes obtained for the Lagrangian with those obtained through 
the double copy.
Since the $g'$-independent terms are the same as in the ungauged theory, we shall focus here on the $g'$-dependent 
amplitudes; we inspect separately the terms quadratic and linear in $g'$.

From a double-copy point of view, the former must contain a four-scalar amplitude in the ${\cal N}=0$ factor. The amplitude
 with four independently-labeled scalars can be found in equation~\eqref{notinpairs} in  a color/kinematics-satisfying form; from here we 
may construct the amplitude with one pair or two pairs of identical scalars.

The ${\cal N}=2$ four-point amplitude labeled in terms of the constrained superfield \eqref{constrainedmultiplets} is
\be
{\cal A}_4^{(0), {\cal N}=2}(1^{\cal G}2^{\cal G}3^{\cal G}4^{\cal G})=\SYMcoupling^2
\left(\frac{{\hat n}_s c_s}{s}+\frac{{\hat n}_t c_t}{t}+\frac{{\hat n}_u c_u}{u}\right)
\,{\cal Q}_4^{34}\,\delta^4(\sum_i\eta_i |i\rangle) \ ,
\ee
with numerator factors obeying the relations
\bea
&& 
\quad
\frac{{\hat n}_s}{s}-\frac{{\hat n}_t}{t}=\frac{i}{\langle 12\rangle\langle 23\rangle\langle 34\rangle\langle 41\rangle}
 \nonumber \ , \\ && \quad
\frac{{\hat n}_t}{t}-\frac{{\hat n}_u}{u}=\frac{i}{\langle 14\rangle\langle 42\rangle\langle 23\rangle\langle 31\rangle}
 \nonumber \ , \\ && \quad
\frac{{\hat n}_u}{u}-\frac{{\hat n}_s}{s}=\frac{i}{\langle 13\rangle\langle 34\rangle\langle 42\rangle\langle 21\rangle} \ ,
\eea
and color factors given by \bea && 
c_s = {\tilde \rf}^{\hat a_1\hat a_2 \hat  b}{\tilde \rf}^{\hat a_3 {\hat a}_4 \hat b} \ ,
\qquad
c_t = {\tilde \rf}^{\hat a_1{\hat a}_4 \hat  b}{\tilde \rf}^{\hat a_2 \hat a_3 \hat  b} \ , 
\qquad
c_u = {\tilde \rf}^{\hat a_1\hat a_3 \hat  b}{\tilde \rf}^{{\hat a}_4 \hat a_2\hat  b} \ .
\eea
It is convenient to choose the kinematic coefficient of one of the color structures to vanish; we will choose ${\hat n}_t=0$.
The kinematics Jacobi identity obeyed by the numerator factors ${\hat n}$ 
implies that ${\hat n}_s=-{\hat n}_u$.
Then, the four-point superamplitude proportional to $(g')^2$ is:
 \be
 \label{4ptgpsq}
 M_4(1^{{\cal V}^{\allphi{ A}_1}}2^{{\cal V}^{\allphi{ A}_2}}3^{{\cal V}^{\allphi{ A}_3}}4^{{\cal V}^{\allphi{ A}_4}})
=  {i \over 2} g'{}^2\,{\cal Q}_4^{34}\delta^{(4)}(\sum_i\eta_i |i\rangle)
\left(
\frac{F^{\allphi{ A}_3\allphi{ A}_1\allphi{ B}}F^{\allphi{ A}_2\allphi{ A}_4\allphi{ B}} }
{\langle 14\rangle\langle 42\rangle\langle 23\rangle\langle 31\rangle} \;
-
\frac{F^{\allphi{ A}_1\allphi{ A}_2\allphi{ B}}F^{\allphi{A}_3\allphi{ A}_4\allphi{ B}}  }
{\langle 12\rangle\langle 23\rangle\langle 34\rangle\langle 41\rangle}\;
                        \right)\ . \\
\ee

The four-point amplitudes of the Yang-Mills-scalar theory have contributions linear in $g'$ corresponding to three scalars and one gluon. It is not hard to see from the Lagrangian  \eqref{vectorscalarL} that, up to permutation of external legs and conjugation, these are given by
\bea
{\cal A}_4^{(0),{\cal N}=0}(1^{\phi^{\allphi{A}_1}}2^{\phi^{\allphi{A}_2}}3^{\phi^{\allphi{A}_3}}4^-)&=&\frac{i \, \SYMcoupling^2 g'}{\sqrt{2}} 
F^{\allphi{A}_1\allphi{A}_2\allphi{A}_3}
\frac{\langle 14\rangle\langle 43\rangle }{\langle 13\rangle}
\left(\frac{1}{s} {\tilde \rf}^{\hat a_1\hat a_2 \hat b}{\tilde \rf}^{\hat a_3 {\hat a}_4 \hat  b}-\frac{1}{t} {\tilde \rf}^{\hat a_1{\hat a}_4 \hat  b}{\tilde \rf}^{\hat a_2 \hat a_3 \hat  b} 
\right) ,
 \\
{\cal A}_4^{(0),{\cal N}=0}(1^{\phi^{\allphi{A}_1}}2^{\phi^{\allphi{A}_2}}3^{\phi^{\allphi{A}_3}}4^+)&=&\frac{i \, \SYMcoupling^2 g'}{\sqrt{2}} 
F^{\allphi{A}_1\allphi{A}_2\allphi{A}_3}
\frac{[ 14][ 43]}{[ 13 ]}
\left(\frac{1}{s} {\tilde \rf}^{\hat a_1\hat a_2 \hat  b}{\tilde \rf}^{\hat a_3 {\hat a}_4 \hat  b}-\frac{1}{t} {\tilde \rf}^{\hat a_1{\hat a}_4 \hat  b}{\tilde \rf}^{\hat a_2 \hat a_3 \hat b} \right)  ,
\qquad \eea
where we picked the reference vector in the gluon polarization vector to be $k_2$.

Then, the resulting supergravity superamplitudes that are linear in the gauge coupling are given by
\bea
\begin{aligned}
M_4(1^{{\cal V}^{\allphi{A}_1}}2^{{\cal V}^{\allphi{A}_2}}3^{{\cal V}^{\allphi{A}_3}}4^{{\cal H}_-})
&=\frac{i}{\sqrt{2}}g'F^{\allphi{A}_1\allphi{A}_2\allphi{A}_3}\; {\cal Q}^{34}_4 
\frac{\delta^4(\sum_i\eta_i |i\rangle)}{\langle 12\rangle\langle 23\rangle\langle 31\rangle}\;
\end{aligned}
\label{4ptgp}
\eea
and the $CPT$-conjugate amplitude.
It is straightforward (albeit quite tedious) to derive the $g'$-dependent terms of four-point amplitudes using standard Feynman 
diagrammatics and see that the maps \eqref{Fmap} and \eqref{multiplet_map} relate them to the double-copy amplitudes listed 
in this section.

We note here that supergravity scattering amplitudes obey color/kinematics duality on all legs for which a Jacobi relation can be 
constructed (and do so manifestly if the ${\cal N}=2$ sYM amplitudes obey the duality). It is not hard to check this assertion, which may be 
understood as a consequence of the color/kinematics duality of the gauge 
theory factors, on 
equations~\eqref{4ptgpsq} and \eqref{4ptgp}. 
Indeed, the internal legs on which Jacobi identities can be constructed are color non-singlets and therefore, from the perspective 
of the double-copy construction, start and end at a trilinear scalar vertex in the Yang-Mills-scalar theory. The part of the numerator 
factors due to these 
vertices is momentum-independent and depends only on the structure constants $F^{\allphi{A}_1\allphi{A}_2\allphi{A}_3}$. 
Thus, whenever the gauge-group color factors obey the Jacobi identity, the global symmetry group factors obey it as well.
In the scattering amplitudes of the corresponding double-copy YMESGT the global symmetry group factors of the Yang-Mills-scalar theory
become color factors and are multiplied by the numerator factors of the ${\cal N}=2$ theory, which are assumed to obey the kinematic 
Jacobi relations. It therefore follows that whenever the YMESGT color factors of an amplitude 
obey Jacobi relations (on a leg on which such a relation 
may be defined) then so do the kinematics numerator factors, {i.e.} the amplitude exhibits manifest color/kinematics duality.

\subsection{Five-point amplitudes}

Having gained confidence that the construction proposed here describes the generic Jordan family of YMESGTs, we can proceed 
to compute higher-point amplitudes.
The double-copy construction of the five-point superamplitudes of YMSGTs is slightly more involved due to the more complicated 
structure of the color/kinematics-satisfying representations of the ${\cal N}=2$ superamplitudes.  Such a representation may be 
obtained as a $\mathbb{Z}_2$ projection of the corresponding ${\cal N}=4$ five-point superamplitude:
\begin{eqnarray}
{\cal A}^{(0)}_5& = & 
\SYMcoupling^3\Bigl( {n_1 c_1\over s_{12}s_{45}}+{n_2 c_2\over s_{23}s_{51}}
+{n_3 c_3 \over s_{34}s_{12}}+{n_4 c_4\over s_{45}s_{23}}
+{n_5 c_5\over s_{51}s_{34}} +{n_{6} c_6\over s_{14}s_{25}}  \nn\\
&&\null \hskip .3 cm 
+ {n_7 c_7\over s_{32}s_{14}}+{n_{8} c_8\over s_{25}s_{43}}
+{n_{9} c_9\over s_{13}s_{25}}+{n_{10} c_{10}\over s_{42}s_{13}}
+{n_{11} c_{11}\over s_{51}s_{42}}+ {n_{12} c_{12}\over s_{12}s_{35}}  \nn\\
&&\null \hskip .3 cm 
+{n_{13} c_{13}\over s_{35}s_{24}} 
+{n_{14} c_{14}\over s_{14}s_{35}}
+{n_{15} c_{15}\over s_{13}s_{45}}\Bigr)\,,
\label{YangMillsNumerator}
\end{eqnarray}
where the color factors are explicitly given by ref.~\cite{BCJ},
\begin{eqnarray}
&& 
c_{1\phantom{0}} \equiv \rff^{\hat a_1\hat a_2 \hat b}\rff^{\hat b \hat a_3\hat  c}\rff^{\hat c {\hat a}_4 {\hat a}_5}\,, \hskip 0.8cm  
c_{2\phantom{1}} \equiv \rff^{\hat a_2 \hat a_3\hat  b}\rff^{\hat b {\hat a}_4 \hat c}\rff^{\hat c {\hat a}_5 \hat a_1}\,, \hskip 0.8cm  
c_{3\phantom{1}} \equiv \rff^{\hat a_3 {\hat a}_4 \hat b}\rff^{\hat b {\hat a}_5 \hat c}\rff^{\hat c \hat a_1\hat a_2}\,, \nn \\&&
c_{4\phantom{1}} \equiv \rff^{{\hat a}_4 {\hat a}_5 \hat b}\rff^{\hat b \hat a_1\hat c}\rff^{\hat c \hat a_2 \hat a_3}\,, \hskip 0.8cm  
c_{5\phantom{1}} \equiv \rff^{{\hat a}_5 \hat a_1\hat b}\rff^{\hat b \hat a_2 \hat c}\rff^{\hat c \hat a_3 {\hat a}_4}\,, \hskip 0.8cm  
c_{6\phantom{1}} \equiv \rff^{\hat a_1{\hat a}_4 \hat b}\rff^{\hat b \hat a_3 \hat c}\rff^{\hat c \hat a_2 {\hat a}_5}\,, \nn \\&& 
c_{7\phantom{1}} \equiv \rff^{\hat a_3 \hat a_2 \hat b}\rff^{\hat b {\hat a}_5 \hat c}\rff^{\hat c \hat a_1{\hat a}_4}\,, \hskip 0.8cm  
c_{8\phantom{1}} \equiv \rff^{\hat a_2 {\hat a}_5 \hat b}\rff^{\hat b \hat a_1\hat c}\rff^{\hat c {\hat a}_4 \hat a_3}\,, \hskip 0.8cm  
c_{9\phantom{1}} \equiv \rff^{\hat a_1\hat a_3 \hat b}\rff^{\hat b {\hat a}_4 \hat c}\rff^{\hat c \hat a_2 {\hat a}_5}\,, \nn \\&&
c_{10} \equiv \rff^{{\hat a}_4 \hat a_2 \hat b}\rff^{\hat b {\hat a}_5 \hat c}\rff^{\hat c \hat a_1\hat a_3}\,, \hskip 0.8cm  
c_{11} \equiv \rff^{{\hat a}_5 \hat a_1\hat b}\rff^{\hat b \hat a_3 \hat c}\rff^{\hat c {\hat a}_4 \hat a_2}\,, \hskip 0.8cm  
c_{12} \equiv \rff^{\hat a_1\hat a_2 \hat b}\rff^{\hat b {\hat a}_4 \hat c}\rff^{\hat c \hat a_3 {\hat a}_5}\,, \nn \\\ &&
c_{13} \equiv \rff^{\hat a_3 {\hat a}_5 \hat b}\rff^{\hat b \hat a_1\hat c}\rff^{\hat c \hat a_2 {\hat a}_4}\,, \hskip 0.8cm  
c_{14} \equiv \rff^{\hat a_1{\hat a}_4 \hat b}\rff^{\hat b \hat a_2 \hat c}\rff^{\hat c \hat a_3 {\hat a}_5}\,, \hskip 0.8cm  
c_{15} \equiv \rff^{\hat a_1\hat a_3 \hat b}\rff^{\hat b \hat a_2 \hat c}\rff^{\hat c {\hat a}_4 {\hat a}_5}\,. \hskip 1.5 cm 
\label{FivePointColor}
\end{eqnarray}
In the ${\cal N}=4$ theory the numerator factors $n_i$ have many different forms, see e.g. ref.~\cite{Broedel:2011pd}
and for each of them the orbifold projection yields an ${\cal N}=2$ superamplitude with the desired properties.
For five-point amplitudes this projection amounts to replacing super-momentum-conserving 
delta function as in equation~\eqref{projection}. An example of numerator factors is
\begin{eqnarray}
\label{rep52}
n(a,b,c,d,e)=
\frac{1}{10} \Bigg(  \left(\frac{1}{s_{c d}}-\frac{1}{s_{c e}}\right) \gamma_{a b} +  \left(\frac{1}{s_{a c}}-\frac{1}{s_{b c}}\right) \gamma_{ed} \no
\\
- \Big[ \frac{\beta_{e d c b a  }}{s_{a e}}+\frac{\beta_{d e c a b  }}{s_{b d}}
-
\frac{\beta_{ e d c a b  }}{s_{b e}}-
\frac{\beta_{d e c  b  a  }}{s_{a d}} \Big] \Bigg)\, ,
\end{eqnarray}
where
\bea
&&
\beta_{12345}\equiv - {\cal Q}_5^{34} \delta^{(4)}(\sum_{i=1}^5\eta_i|i\rangle ) 
\frac{  [12] [23][34][45][51] }{4 \, \varepsilon(1,2,3,4)} \no \ ,
\\
&&\quad\;
\gamma_{ij}\equiv \gamma_{ijklm} = \beta_{ijklm}-\beta_{jiklm} \ .
\eea
The order of arguments is given by the order of free indices of the color factor.

The ${\cal N}=0$ five-scalar amplitude has the same form as \eqref{YangMillsNumerator} except that the numerator factors 
are (quadratic) polynomials in $g'$:
\be
n_i = n_{i,0}+ \frac{1}{2} g' F^{{a}_1{a}_2{a}_3}  n_{i,1}+g'{}^3 
F^{{a}_1{a}_3{b}} F^{{b}{a}_3{c}} F^{{c}{a}_3{a}_2} n_{i,2} 
\ee
with $n_{i,0}=0$ for all $i=1,\dots,15$. Eqs.~\eqref{5Sgp3} and \eqref{5Sgp1} determine $n_{i,1}$ and $n_{i,2}$ to be\footnote{We use the notation $k_{12{\bar 3}} = k_1+k_2-k_3$, { etc}.}
\bea
\begin{array}{lclclclcl}
n_{1,1} = k_{12{\bar 3}}\cdot k_{4{\bar 5}}
&&
n_{2, 1}=-(s_{15}+s_{23})
&&
n_{3, 1}=-k_{12{\bar 5}}\cdot k_{3{\bar 4}}
\\
n_{4, 1}={k_{{\bar 1}23}\cdot k_{4{\bar 5}}}
&&
n_{5, 1}=-{k_{1{\bar 2}5}\cdot k_{3{\bar 4}}}
&&
n_{6, 1}=-({s_{14}+s_{25}})
\\
n_{7, 1}=-({s_{14}+s_{23}})
&&
n_{8, 1}=-{k_{{\bar 1}25}\cdot k_{3{\bar 4}}}
&&
n_{9, 1}=-({s_{13}+s_{25}})
\\
n_{10, 1}=-({s_{13}+s_{24}})
&&
n_{11, 1}=-({s_{15}+s_{24}})
&&
n_{12, 1}={k_{12{\bar 4}}\cdot k_{3{\bar 5}}}
\\
n_{13, 1}=-{k_{{\bar 1}24}\cdot k_{3{\bar 5}}}
&&
n_{14, 1}=-{k_{1{\bar 2}4}\cdot k_{3{\bar 5}}}
&&
n_{15, 1}=-{k_{1{\bar 2}3}\cdot k_{4{\bar 5}}}
\end{array}
\eea
and
\bea
\begin{array}{lclclclcl}
n_{1, 2} = 0
&&
n_{2, 2}=-1
&&
n_{3, 2}=0
&&
n_{4, 2}=0
&&
n_{5, 2}=0
\\
n_{6, 2}=-1
&&
n_{7, 2}=-1
&&
n_{8, 2}=0
&&
n_{9, 2}=-1
&&
n_{10, 2}=-1
\\
n_{11, 2}=-1
&&
n_{12, 2}=0
&&
n_{13, 2}=0
&&
n_{14, 2}=0
&&
n_{15, 2}=0
\end{array}
\eea
For a more general choice of scalars than considered here all 15 numerator factors are nonzero.
The ${\cal O}(g'{}^3)$ part of the corresponding five-vector superamplitude is given by
\bea
&&
{\cal M}_5^{(0)}(1^{{\cal V}^{\allphi{A}_1}}2^{{\cal V}^{\allphi{A}_2}}3^{{\cal V}^{\allphi{A}_3}}4^{{\cal V}^{\allphi{A}_3}}
5^{{\cal V}^{\allphi{A}_3}})|_{(g')^3}=
-i\left(\frac{\kappa}{2}\right)^3 (g')^3F^{{  \allphi{A}}_1{  \allphi{A}}_3{  \allphi{B}}} F^{{  \allphi{B}}{  \allphi{A}}_3{  \allphi{C}}} 
F^{{  \allphi{C}}{  \allphi{A}}_3{  \allphi{A}}_2}
\\
&&\qquad\qquad\qquad\qquad
\times
\Bigl({n_2 \over s_{23}s_{51}}
      +{n_{6} \over s_{14}s_{25}} 
+ {n_7  \over s_{32}s_{14}}
+{n_{9}  \over s_{13}s_{25}}+{n_{10} \over s_{42}s_{13}}
+{n_{11}  \over s_{51}s_{42}}+ {n_{12} \over s_{12}s_{35}}  
\Bigr) \ .
\nonumber
\eea
We notice that the numerator factors are simply those of the ${\cal N}=2$ sYM theory and thus they obey the Jacobi relations
simultaneously with the $F$ color factors.

The ${\cal O}(g')$ part of the five-vector superamplitude with three different gauge indices is given by
\bea
&&
{\cal M}_5^{(0)}(1^{{\cal V}^{\allphi{A}_1}}2^{{\cal V}^{\allphi{A}_2}}3^{{\cal V}^{\allphi{A}_3}}
4^{{\cal V}^{\allphi{A}_3}}5^{{\cal V}^{\allphi{A}_3}})|_{g'}=
\frac{i}{2} \left(\frac{\kappa}{2}\right)^3 g' F^{{  \allphi{A}}_1{  \allphi{A}}_2{  \allphi{A}}_3} \no
\\
&&\qquad\qquad\qquad
\times \Bigl( {n_1 n_{1, 1} \over s_{12}s_{45}}+{n_2 n_{2, 1} \over s_{23}s_{51}}
+{n_3 n_{3 , 1} \over s_{34}s_{12}}+{n_4 n_{4, 1} \over s_{45}s_{23}}
+{n_5 n_{5, 1} \over s_{51}s_{34}} +{n_{6} n_{6, 1} \over s_{14}s_{25}}  \nn\\
&&\qquad\qquad\qquad\null \hskip .3 cm 
+ {n_7 n_{7, 1} \over s_{32}s_{14}}+{n_{8} n_{8, 1} \over s_{25}s_{43}}
+{n_{9} n_{9, 1} \over s_{13}s_{25}}+{n_{10} n_{{10}, 1} \over s_{42}s_{13}}
+{n_{11} n_{{11}, 1} \over s_{51}s_{42}}+ {n_{12} n_{{12}, 1} \over s_{12}s_{35}}  \nn\\
&&\qquad\qquad\qquad\null \hskip .3 cm 
+{n_{13} n_{{13}, 1} \over s_{35}s_{24}} 
+{n_{14} n_{{14}, 1} \over s_{14}s_{35}}
+{n_{15} n_{{15}, 1} \over s_{13}s_{45}}\Bigr)
\ .
\eea
By undoing the projection \eqref{projection} one recovers the five-point amplitude of ${\cal N}=4$ supergravity coupled with abelian and non-abelian vector multiplets.

Using similar higher-point color/kinematics-satisfying representations of tree-level ${\cal N}=2$ amplitudes, 
perhaps constructed in terms of color-ordered amplitudes or by some other methods, and Feynman-graph generated
amplitudes of the Yang-Mills-scalar theory, it is easy to construct tree-level amplitudes of any multiplicity for 
YMESGTs in the generic Jordan family. 

The analysis described in this section can also be carried out in five dimensions with similar conclusions: a five-dimensional 
YMESGT in the generic Jordan family can be described as a double-copy of the half-maximal sYM theory and a vector-scalar 
theory with trilinear couplings. On the double-copy side there is a trivial map between the calculation described in this section and
its five-dimensional version given by (inverse) dimensional reduction. 
Since five-dimensional MESGTs and YMESGTs are uniquely specified by their trilinear couplings, this suggests 
that our identification of the supergravity theory is unique. At the level of the supergravity Lagrangian this may not be obvious 
because on the one hand we use the $SO(\tilde n)$ symplectic frame to carry out the discussion in this section while on the 
other hand the lift to five dimensions is natural in a different frame. At tree level one should simply invert the transformations discussed 
in sec.~\ref{sec4dsymp}. At loop level the presence of duality anomalies might prevent this program from being successful. 
One may however carry out this program at the level of the dimensionally-regularized integrands of loop amplitudes which are, by 
construction, the quantities that have a  smooth continuation in dimension. From this perspective, the loop integrands of the scattering 
amplitudes relevant for the four-dimensional duality anomaly are part of the integrands of typical five-dimensional 
amplitudes (e.g. the $4D$ two-graviton-two-holomorphic-scalar amplitude is part of the $5D$ four-graviton amplitude); the anomaly 
manifests itself as such only after the loop momentum integral is evaluated close to four dimensions.  
We therefore conclude that the four-dimensional analysis described in this section uniquely singles out the YMESGTs as being 
given by the double-copy construction using ${\cal N}=2$ sYM and a vector-scalar theory with trilinear couplings.


\section{One-loop four-point amplitudes \label{oneloop}}

Similarly to tree amplitudes, loop amplitudes in YMESGTs can be organized following the dependence on the gauge coupling;
each component with a different gauge coupling factor is separately gauge invariant.
For the YMESGTs considered here it is not difficult to argue both from a Lagrangian and double-copy point of view that, to any 
loop order and multiplicity, the terms with the highest power of the gauge coupling in the  $n$-vector amplitudes are given by the 
amplitudes of a pure ${\cal N}=2$ sYM theory with the same gauge group as that of the supergravity theory.\footnote{The same 
assertion holds in ${\cal N}=4,1,0$ YMESGTs.}
From a double-copy perspective these terms are given by the amplitudes of the Yang-Mills-scalar theory with only scalar 
vertices; since these amplitudes have constant numerator factors, when double-copied with the amplitudes of ${\cal N}=2$ 
or ${\cal N}=4$ sYM theory (or any other theory for that matter), they simply replace the color factors of the latter with the 
color factors of the supergravity gauge group.
From a Lagrangian perspective the terms with highest power of the gauge coupling in the vector amplitudes are given by the 
$\kappa\rightarrow 0$ limit of the full amplitude and thus are given by a pure gauge-theory computation.

To illustrate the construction of loop amplitudes in the generic Jordan family of YMESGTs we shall compute the 
simplest  one-loop amplitude that is sensitive to the supergravity gauge coupling---the four-vector amplitude.
To this end we will first find the bosonic Yang-Mills-scalar amplitude with external scalar matter in a form that manifestly obeys 
color/kinematics duality. Then, through the double copy, this amplitude will be promoted to be a four-vector amplitude in ${\cal N}=4$ 
and ${\cal N}=2$ YMESGTs.

\subsection{The four-scalar gauge-theory amplitude}

\begin{figure}[t]
      \centering
      \includegraphics[scale=0.9]{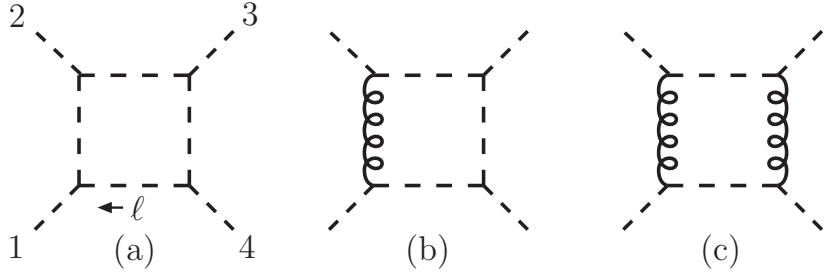}
      \caption[a]{\small The three types of diagrams that contribute at different orders in the $g'$ coupling to
      the box numerator of a four-scalar one-loop amplitude. All distinct cyclic permutations of these diagrams should be included. 
      Dashed lines denote scalar fields while curly lines denote vector fields. Note that quartic-scalar interactions are implicitly included in these diagrams, according to their power in the $g'$ coupling.
               \label{OneLoopYMFigure} }
\end{figure}

The three classes of Feynman graphs contributing to the ${\cal O}(g'{}^4)$, ${\cal O}(g'{}^2)$ and ${\cal O}(g'{}^0)$ terms in the 
four-scalar amplitude of the Yang-Mills-scalar theory~\eqref{YMStheory} are schematically shown in \fig{OneLoopYMFigure}.
The ${\cal O}(g'{}^4)$ is the simplest as it  fully correspond to the four-scalar amplitude in the $\phi^3$ theory.
The numerator of the box diagram, shown in \fig{OneLoopYMFigure}(a), is entirely expressed in terms of the 
structure constants of the global group,
\be
n^{\rm(a)}_{\rm box}(1,2,3,4)=-i g'{}^4 F^{b   a_1  c} F^{  c a_2  d} 
F^{  d a_3  e} F^{  e{ a}_4 b} \ ,
 \label{numA}
\ee
where we stripped off the color factor of the gauge group given by 
$c^{\rm(a)}_{\rm box}= \rf^{\hat b {\hat a}_1 \hat c} \rf^{\hat c{\hat a}_2 \hat d} 
\rf^{\hat d{\hat a}_3\hat e} \rf^{\hat e{\hat a}_4\hat b}$.

Next we consider the ${\cal O}(g'{}^2)$ contributions, which correspond to mixed interactions in the Yang-Mills-scalar theory. 
For the box diagram, these contributions are given by \fig{OneLoopYMFigure}(b) and its cyclic permutations. 
While a good approximation for the duality-satisfying box numerator can be obtained using the Feynman rules that follow 
from the Lagrangian \eqref{YMStheory}, we construct the full numerator using an Ansatz constrained to give the
correct unitarity cuts. In the labeling convention of \fig{OneLoopYMFigure}, this gives the following numerator:
\bea
n^{\rm(b)}_{\rm box}(1,2,3,4,\ell)\!&=&\!
-\frac{i}{12} g'^2 \Big\{ (N_V + 2) \big(F^{ a_1{ a}_4   b} F^{  b a_3 a_2}  (\ell_2^2 + \ell_4^2) + F^{  a_1{ a}_2   b} F^{  b a_3 a_4} (\ell_1^2 + \ell_3^2)\big) \nn \\ && \null +24 \big(s  F^{  a_1{ a}_4   b} F^{  b a_3 a_2}  +t F^{  a_1{ a}_2   b} F^{  b a_3 a_4} \big) +
   \delta^{a_3 a_4} \Tr_{12}  (6 \ell_3^2 - \ell_2^2 - \ell_4^2)  \label{numB}  \\ &&
  \null  + \delta^{a_2 a_3} \Tr_{14} (6 \ell_2^2 - \ell_1^2 - \ell_3^2)  +
      \delta^{a_1 a_4} \Tr_{23} (6 \ell_4^2 - \ell_1^2 - \ell_3^2) \nn \\ &&
      \null + \delta^{a_1 a_2} \Tr_{34} (6 \ell_1^2 - \ell_2^2 - \ell_4^2)  
      + (\ell_1^2 + \ell_2^2 + \ell_3^2 + \ell_4^2) (\delta^{a_2 a_4} \Tr_{13} + \delta^{a_1 a_3} \Tr_{24}) \Big\}\,,  \no 
 \eea
where $\ell_j = \ell-(k_1+\ldots+k_j)$, and we use the shorthand notation ${\rm Tr}_{ij}=  F^{b  a_i c} F^{c{a}_j b}$. The parameter $N_V=\delta^{ab}\delta_{ab}$ is the number of scalars in the four-dimensional theory (or the number of vectors after double-copying it with another YM numerator).  In $D$ dimensions, one should replace $N_V \rightarrow N_V + D - 4$ for a consistent state counting.  Note that the box numerator is designed to satisfy the following automorphism identities:  $n^{\rm (b)}_{\rm box}(1,2,3,4,\ell)=n^{\rm(b)}_{\rm box}(4,3,2,1,-\ell)=n^{\rm(b)}_{\rm box}(2,3,4,1,\ell-k_1)$.

Finally, the ${\cal O}(g'{}^0)$ contributions, illustrated by \fig{OneLoopYMFigure}(c) and its cyclic permutations, correspond 
to YM interactions at all vertices. Using the same procedure, an Ansatz constrained by the unitarity cuts of the theory \eqref{YMStheory},
the resulting  duality-satisfying box numerator is given by
\bea
n^{\rm (c)}_{\rm box}(1,2,3,4,\ell)\!&=&\!-\frac{i}{24} \Big\{  \delta^{ a_1 a_2}  \delta^{ a_3 a_4}  \big[24 t (t - 2 \ell_1^2 - 2 \ell_3^2)  +2 (N_V + 2) (3 \ell_1^2 \ell_3^2 - \ell_2^2 \ell_4^2) \nn \\ && 
\hskip 2.4cm \null  + (N_V + 18) \big(t (\ell_1^2 + \ell_2^2 + \ell_3^2 + \ell_4^2) - u (\ell_1^2 + \ell_3^2)\big) \big]  \nn \\ &&
      \null ~ \,\,+ \delta^{ a_2 a_3} \delta^{ a_1 a_4}   \big[24 s (s - 2 \ell_2^2 - 2 \ell_4^2)+2 (N_V + 2) (3 \ell_2^2 \ell_4^2 - \ell_1^2 \ell_3^2)   \nn \\ &&
\hskip 2.4cm \null      + (N_V + 18) \big(s (\ell_1^2 + \ell_2^2 + \ell_3^2 + \ell_4^2) - u (\ell_2^2 + \ell_4^2)\big)\big]  \nn \\ &&
     \null ~ \,\,+ \delta^{ a_1 a_3} \delta^{ a_2 a_4}  \big[2 (N_V + 2) (\ell_1^2 \ell_3^2 + \ell_2^2 \ell_4^2)  \nn \\ && 
  \hskip 2.4cm   \null - (N_V + 18) \big(s (\ell_1^2 + \ell_3^2) + t (\ell_2^2 + \ell_4^2)\big)\big]\Big\}\,, 
\label{numC}
\eea
where  $\ell_i$, $N_V$ and the automorphism identities are the same as for the ${\cal O}(g'{}^2)$ numerator.  
Certain terms in (\ref{numB}) and (\ref{numC}) can be directly identified as contributions from the Feynman rules for the box diagram; 
however, most terms have different origin. They are moved into the box numerator from bubble and triangle graphs by a generalized gauge 
transformation~\cite{BCJ,BCJLoop}. 
This explains the presence of global-group invariants naively not associated with the box diagram.

The box numerators \eqref{numA}, \eqref{numB}, \eqref{numC} are constructed so that the amplitude manifestly obeys color/kinematics duality. 
In particular, the numerator factors for the remaining contributing diagrams, the triangles and bubbles, are given by the color-kinematical 
Lie algebra relations:
\bea
n^{(x)}_{\rm tri.}([1,2],3,4,\ell)&=&n^{(x)}_{\rm box}(1,2,3,4,\ell)-n^{(x)}_{\rm box}(2,1,3,4,\ell)\,, \nn \\
n^{(x)}_{\rm bub.}([1,2],[3,4],\ell)&=&n_{\rm tri.}^{(x)}([1,2],3,4,\ell)-n_{\rm tri.}^{(x)}([1,2],4,3,\ell) \ ,
\eea
where the notation $[i,j]$  implies that external legs $i$ and $j$ meet at the same vertex.
Similarly, the gauge-group color factors for all box diagrams are given by
\be
c_{\rm box}=c^{\rm(a)}_{\rm box}=c^{\rm(b)}_{\rm box}=c^{\rm(c)}_{\rm box}
= \rf^{\hat b\hat a_1\hat c} \rf^{\hat c\hat a_2\hat d} \rf^{\hat d\hat a_3\hat e} \rf^{\hat e{\hat a}_4\hat b}\ ,
\ee
and the remaining color factors are determined by the Jacobi relations of the gauge-group Lie algebra,
\bea
c_{\rm tri.}([1,2],3,4)&=&c_{\rm box}(1,2,3,4)-c_{\rm box}(2,1,3,4) \ , \nn \\
c_{\rm bub.}([1,2],[3,4])&=&c_{\rm tri.}([1,2],3,4)-c_{\rm tri.}([1,2],4,3)\ .
\eea

The complete numerators are given by the sum of the ${\cal O}(g'{}^4)$, ${\cal O}(g'{}^2)$, ${\cal O}(g'{}^0)$ contributions,
\be
n_i=n^{\rm (a)}_i+n^{\rm (b)}_i+n^{\rm (c)}_i\,.
\label{fullNum}
\ee
We have verified that these numerators are correct on all unitarity cuts in $D$ dimensions;
in particular non-planar single-particle cuts $\sum_{\rm states } A_6^{(0)}(1,2, \ell, 3,4, -\ell)$ were used in this check.

\subsection{The four-vector Yang-Mills--gravity amplitude}

\begin{figure}[t]
      \centering
      \includegraphics[scale=0.8]{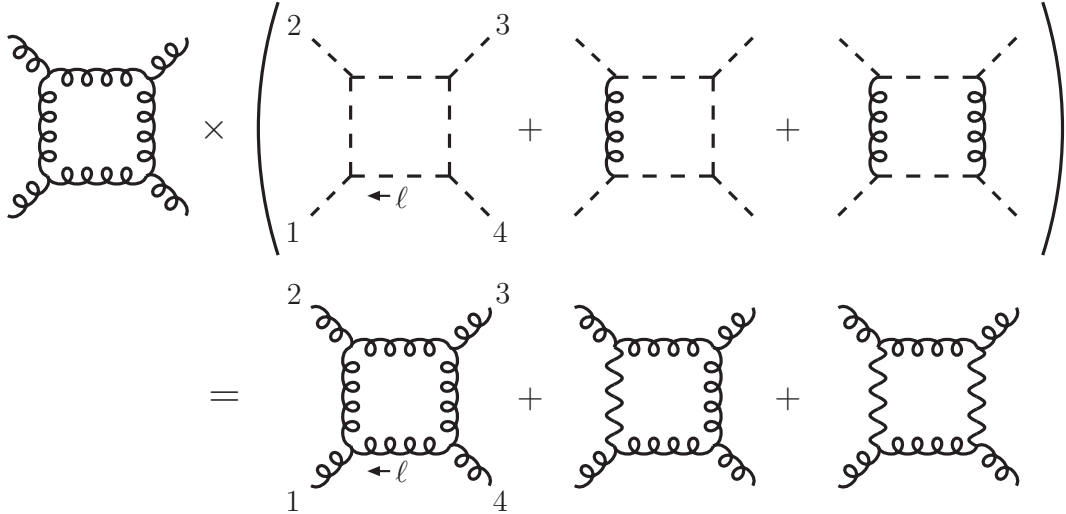}
      \caption[a]{\small The three types of diagrams that contribute to the box numerator in YMESG can be obtained through a double copy between the diagrams in \fig{OneLoopYMFigure} and a sYM numerator that obeys color/kinematics duality. All distinct cyclic permutations of these diagrams should be included. We use dashed lines to denote 
      scalar fields, curly lines to denote vector fields or vector multiplets (as the case may be) and wavy lines  to 
      denote the graviton multiplet.
               \label{OneLoopGRFigure}}
\end{figure}

The double-copy recipe provides a straightforward way to construct amplitudes in the ${\cal N}=2$ 
of the generic Jordan family as well as ${\cal N}=4$ YMESGTs. For example, \fig{OneLoopGRFigure} 
illustrates how the different types of contributions ${\cal O}(g^4)$, ${\cal O}(g^2)$, ${\cal O}(g^0)$ arises as 
double copies between sYM numerators and numerators computed in the previous section.
As in the case of other supergravity theories, one 
may verify that the unitarity cuts of these amplitudes match the direct evaluation of cuts in terms of tree diagrams.

The complete amplitude is given by the double-copy formula \eqref{DoubleCopy},
\begin{equation}
 {\cal M}^{1-\text{loop}}_4 = - \, \left(\frac{\kappa}{2}\right)^{4} \,
\sum_{{\cal S}_4}\sum_{i\in{\rm\{box,tri.,bub.\}}} {\int{  \frac{d^D \ell}{(2 \pi)^D}
 \frac{1}{S_i}
   \frac{n_i {\tilde n}_i}{D_i}}} \,,
\hskip .7 cm 
\label{1LoopDC}
\end{equation}
where the first sum runs over the permutations ${\cal S}_4$ of all four external leg labels; the symmetry factors 
are $S_{\rm box}=8$, $S_{\rm tri.}=4$ and $S_{\rm bub.}=16$. The denominator factors, $D_i$, are
\be
D_{\rm box}=\ell_1^2\ell_2^2\ell_3^2\ell_4^2\,,~~D_{\rm tri.}=s \ell_2^2\ell_3^2\ell_4^2\,,~~~D_{\rm bub.}=s^2 \ell_2^2\ell_4^2\,,
\ee
where $\ell_i=\ell-(k_1+\ldots+k_i)$. The $n_i$ are given in \eqref{fullNum}, and the ${\tilde n}_i$ are numerators of sYM theories. Moreover, as at tree level, we identify the $g'$ parameter of the Yang-Mills-scalar theory 
with the supergravity gauge coupling $g$.

Standard on-shell supersymmetry arguments imply that ${\cal N}=2$ one-loop numerator factors may be written as
the difference between ${\cal N}=4$ and numerator factors for one adjoint ${\cal N}=2$ hypermultiplet running in the loop, 
\be
{\tilde n}^{{\cal N}=2}_{i}(1,2,3,4)={\tilde n}^{{\cal N}=4}_{i}(1,2,3,4)-2  {\tilde n}_{i}^{{\cal N}=2,{\rm mat.}}(1,2,3,4,\ell) \ .
\ee
The ${\cal N}=4$ sYM box numerator is given by
\be
\tilde{n}^{{\cal N}=4}_{\rm box}(1,2,3,4,\ell)= s t A^{\tree}(1,2,3,4) = -i\frac{\spb{1}.{2}\spb{3}.{4}}{\spa{1}.{2}\spa{3}.{4}}\delta^{(4)}(\sum \eta_i^\alpha |i\rangle)\ ,
\ee
and the triangle and bubble numerator vanishes. Plugging this into \eqref{1LoopDC} gives  the four-vector amplitude in 
${\cal N}=4$ YMESGT.

Color/kinematics-satisfying one-loop numerator factors due to one adjoint hypermultiplet running in the loop may be found in refs.~\cite{Carrasco:2012ca,Nohle:2013bfa,Ochirov:2013xba,Johansson:2014zca}. A manifestly ${\cal N}=2$-supersymmetric 
box numerator was given in ref.~\cite{Johansson:2014zca},
\bea
{\tilde n}_{\rm box}^{{\cal N}=2,{\rm mat.}}(1,2,3,4,\ell)
      &=& \, (\kappa_{12}+\kappa_{34}) \frac{(s-\ell_s)^2}{2s^2}
        + (\kappa_{23}+\kappa_{14}) \frac{\ell_t^2}{2t^2}
        + (\kappa_{13}+\kappa_{24}) \frac{st+(s+\ell_u)^2}{2u^2} \nn \\
    &&\null + \mu^2 \Big( \frac{\kappa_{12}+\kappa_{34}}{s}
                     +\frac{\kappa_{23}+\kappa_{14}}{t}
                     +\frac{\kappa_{13}+\kappa_{24}}{u} \Big) \nn\\
      &&
      \null -2i \epsilon(1,2,3,\ell)\frac{\kappa_{13}-\kappa_{24}}{u^2} \ ,\label{Neq2matter}
\eea
where $\ell_s=2 \ell \cdot (k_1+k_2)$, $\ell_t=2 \ell \cdot (k_2+k_3)$ and $\ell_u=2 \ell \cdot (k_1+k_3)$. The numerator 
factors of other box integrals are obtained by relabeling. The parameter $\mu$ is the component of the loop momenta that 
is orthogonal to four-dimensional spacetime. The external multiplet dependence is captured by the variables $\kappa_{ij}$, 
\be
   \kappa_{ij} = i\frac{ \spb{1}.{2} \spb{3}.{4} }{ \spa{1}.{2}\!\spa{3}.{4} }
   \delta^{(4)}(\sum \eta_i^\alpha |i\rangle) \spa{i}.{j}^{2} (\eta^3_i \eta^4_i)(  \eta^3_j \eta^4_j)   \ .
\label{kappa}
\ee
As before the triangle and bubble numerators are given by the kinematic Jacobi relations,
\bea
{\tilde n}^{{\cal N}=2}_{\rm tri.}([1,2],3,4,\ell)&=&{\tilde n}^{{\cal N}=2}_{\rm box}(1,2,3,4,\ell)
-{\tilde n}^{{\cal N}=2}_{\rm box}(2,1,3,4,\ell)\,, \nn \\
{\tilde n}^{{\cal N}=2}_{\rm bub.}([1,2],[3,4],\ell)&=&{\tilde n}^{{\cal N}=2}_{\rm tri.}([1,2],3,4,\ell)
-{\tilde n}^{{\cal N}=2}_{\rm tri.}([1,2],4,3,\ell) \ .
\eea
Plugging these numerators together with the Yang-Mills-scalar numerators \eqref{fullNum} into equation~\eqref{1LoopDC} gives  
the four-vector amplitude in the ${\cal N}=2$ YMESGT.
The resulting expression exhibits some of the properties outlined in the beginning of section~\ref{oneloop} and at the end 
of section~\ref{fourpttree}. We notice, in particular, that the ${\cal O}(g'{}^4)$ terms are given entirely by the ${\cal N}=2$ numerator 
factors while the scalar amplitudes in the Yang-Mills-scalar theory provide only the color factors of the supergravity gauge group.
As such, these terms are precisely an ${\cal N}=2$ sYM amplitude and, for our choice of numerator factors, manifestly obeys 
color/kinematics duality.

Part of the divergence structure of the amplitude may be inferred from these properties. The ${\cal O}(g'{}^0)$, being the four-vector 
amplitude of the corresponding generic Jordan family MESGT, exhibits a one-loop divergence proportional to the square of the vector 
field stress tensor \cite{Fischler:1979yk, Fradkin:1983xs}. Similarly, the ${\cal O}(g'{}^4)$ is the four-gluon amplitude of pure  ${\cal N}=2$ 
sYM theory and therefore diverges in four dimensions. Finding the ${\cal O}(g'{}^2)$ require the evaluation of the integrals; carrying this 
out leads to a finite result in four dimensions and shows that gauging does not worsen the UV behavior of the ungauged theory. 
It would be interesting to understand how supergravity divergences are affected by the inclusion of the counterterms renormalizing the 
two (s)YM theories.

Although we will not give the details here, we note that one can easily further generalize this calculation to less supersymmetric theories. 
In particular, the duality-satisfying four-point one-loop numerators of ${\cal N}=0$ YM and ${\cal N}=1$ sYM, 
given in refs.~\cite{Carrasco:2012ca, Bern:2013yya, Nohle:2013bfa, Ochirov:2013xba,Johansson:2014zca}, can be inserted into~\eqref{1LoopDC}, 
after which one obtains vector amplitudes in certain ${\cal N}=0$ and ${\cal N}=1$ truncations of the generic Jordan family of YMESGTs.

\newpage
\section{Conclusions and outlook \label{conclusions}}

It is no surprise that MESGTs obtained directly from ${\cal N}=8$ supergravity have a double-copy structure inherited from that 
of the parent theory. Such theories include ${\cal N}=2$ supergravities with $1$, $3$, $5$, or $7$ 
vector multiplets in four dimensions, which can be seen as field-theory orbifolds of maximal supergravity 
and have been studied in \cite{Carrasco:2012ca}.  
It is however less straightforward that the double-copy structure can be extended to theories that have a richer matter content; 
a large class of examples is provided by theories of the generic Jordan family of ${\cal N}=2$  MESGTs, which have particular symmetric target spaces. 

In this paper we studied the formalism that follows from the requirement that supergravity 
theories coupled to non-abelian gauge fields have a double-copy structure. 
We found that gauging global symmetries of ${\cal N}=2$ MESGTs may be accomplished in this framework by adding a certain relevant trilinear scalar operator
to one of the two gauge-theory double-copy factors. The appropriate undeformed theory is a bosonic Yang-Mills-scalar theory that contains quartic scalar interactions consistent with a higher-dimensional pure-YM interpretation. The undeformed gauge theory naturally obeys color/kinematics duality, and we have shown that the deformed gauge theory continues to obey the duality provided that the tensors controlling the trilinear couplings obey 
Jacobi relations and hence can be identified as the structure constants of a gauge group. 
The fact that the gauge theories on both sides of the double copy satisfy the BCJ amplitude relations gives confidence that the construction should give gravitational amplitudes belonging to some well-defined class of supergravity theories.

We discussed in detail the theories in the generic Jordan 
family of ${\cal N}=2$  MESGTs and YMESGTs, and constructed some simple examples of tree-level and one-loop  scattering amplitudes.
By comparing the tree-level result of the double-copy construction and that of a Feynman-graph calculation 
we identified a linearized transformation that relates the Lagrangian and double-copy asymptotic states. 
In particular, for the specific Lagrangian chosen in section~\ref{sec4dsymp}, the two sets of states are related by 
the identity map. Thus we have shown that the specific double-copy construction discussed here gives the 
generic Jordan family of ${\cal N}=2$  MESGTs and YMESGTs in $D=4$ and $D=5$.

Quite generally, there are many possible choices of 
fields that are classically equivalent. As discussed in section~\ref{sec4dsymp}, different choices are related by 
a change of symplectic section, {i.e.} by transformations that include electric/magnetic duality transformations.  
The double-copy realization of electric/magnetic duality transformations  in supergravity was discussed 
in ref.~\cite{Carrasco:2013ypa}, where the charges of supergravity fields were identified as the difference of helicities 
in the two gauge-theory factors,  whenever the  corresponding $U(1)$ transformation is not part of the on-shell 
$R$-symmetry.
It was also shown in ref.~\cite{Carrasco:2013ypa} that, while electric/magnetic duality is a tree-level symmetry 
(tree-level scattering amplitudes carrying nonzero charge vanish), for ${\cal N}\le 4$ theories it acquires a quantum 
anomaly and certain one-loop scattering amplitudes carrying non-vanishing charge are nonzero. 
In the current work we implicitly assume that dimensional regularization is used a loop level, and in that case the anomaly 
appears because the electric/magnetic duality does not lift smoothly between 
spacetime dimensions.\footnote{In principle, 
different regularization schemes may lead to different manifestations of this anomaly. In ref~\cite{Carrasco:2013ypa} as well 
as here dimensional regularization was assumed, which does not preserve electric/magnetic duality.} 
Some amplitudes breaking this duality are the same in the MESGT and in the corresponding YMESGT;  an example 
is the amplitudes with two positive-helicity 
gravitons and two scalars in the ${\widetilde{\rm V}}_-$ multiplet  \cite{Carrasco:2013ypa}. In dimensional regularization it
is 
\be
{\cal M}_4^{(1)}(h_1^{++}, h_2^{++},S^{+-}_3, S^{+-}_4)=\frac{i}{2(4\pi)^2}({\tilde n}+2)
\left(\frac{\kappa}{2}\right)^4[12]^4
\Big[1+2\Big(1-\frac{tu}{3s^2}\Big)\Big] \ ,
\ee
where $S^{+-}\equiv S_{-+}$ and ${\tilde n}$ is the number of vector multiplets. In the presence of such an anomaly
MESGTs and YMESGTs with symplectic sections related classically by duality transformations are no longer equivalent 
quantum mechanically and their effective actions differ by finite local terms in a scheme in which the anomaly is one-loop exact (the existence of 
such a scheme is a consequence of the Adler-Bardeen theorem; see \cite{Carrasco:2013ypa} for a more complete discussion). 
The field identification found in section~\ref{3pf_field_parameter_map} shows that the double-copy construction discussed in this paper realizes a specific symplectic section (through dimensional regularization this gives specific quantum corrections). It would be interesting to see if it is possible to give double-copy constructions for different symplectic sections of the same theory.  

Four-dimensional YMESGTs of the generic Jordan family coupled to hypermultiplets appear as low-energy effective theories 
of the heterotic string compactified on  $K3\times T^2$. 
The string-theory construction suggests that it should be possible to extend our construction 
to include hypermultiplets and their interactions;  tree-level KLT-like relations should exist at least for the 
specific numbers of vector and hypermultiplets that can be accommodated in string theory (or even beyond that using the formalism of ref.~\cite{Johansson:2014zca}).
In particular, it would be desirable to understand how the introduction of hypermultiplets modifies the 
two gauge-theory factors and what are the restrictions on the gauge-group representations 
imposed by coupling ${\cal N}=2$ Yang-Mills-matter theories to ${\cal N}=2$ supergravity.

An extension of the double-copy construction to gauge-theory factors with fields in fundamental and 
bifundamental representations  was discussed in refs.~\cite{Chiodaroli:2013upa, Johansson:2014zca}. 
Generically this yields supergravity theories with different matter content than a double-copy construction with fields in the adjoint representation. 
A natural direction for further research would be to include in the two gauge-theory factors fields transforming in arbitrary representations of the gauge group, and 
to systematically study the gauged supergravity theories obtained with the double copy. 
In particular, a construction of this sort may be necessary to obtain some of the magical supergravity theories 
and to study their gaugings.

In addition to the more formal discussion, we have in this paper also obtained simple expressions for the three-point supergravity superamplitudes. The structure of these 
amplitudes should extend to more general ${\cal N}=2$ MESGTs and YMESGTs, which do not belong to the generic Jordan 
family. The on-shell three-point interactions are universal except for the $C$ tensor, which is used to specify the theory. 
Using this structure it should be possible to construct amplitudes in more general theories from simple building blocks 
even when a double-copy construction is not yet available. 

Understanding whether it is possible to satisfy the locality and dimensionality constraints from having 
a double-copy construction of $R$-symmetry gaugings, as discussed in section~\ref{minimalsec}, 
remains an interesting open problem. 
If such  structure exists, it would be interesting to explore whether it is restricted to 
scattering amplitudes around Minkowski vacua or if it exists, with appropriate choice of boundary conditions, in the 
more general case of AdS vacua.  
This double-copy structure may translate, through the AdS/CFT correspondence, to a double-copy structure 
for the correlation functions of certain gauge-invariant operators of the dual gauge theory. A direct investigation of 
double-copy properties of correlation functions of gauge theories, perhaps along the lines 
of refs.~\cite{Broedel:2012rc, Engelund:2012re, Boels:2012ew}, may also provide an alternative approach to answering  this question.

As mentioned in the beginning of section~\ref{CKsec}, directly applying our construction in six dimensions yields a theory 
that, apart from the graviton multiplet, contains one self-dual tensor and ${\tilde n}-2$ vector multiplets. %
The spectrum of our six-dimensional double-copy construction coincides with that of the $D=6$ YMESGT formulated 
in refs.~\cite{Nishino:1984gk,Nishino:1997ff}. It would be interesting to explore whether the scattering amplitudes following from the Lagrangian of this $D=6$ YMESGT 
are the same as those generated by our construction. Such YMESGTs coupled to hypermultiplets arise from compactification of the heterotic string on  a $K3$ surface. 
%
As remarked earlier the $D=5$ generic Jordan  family  of MESGTs can also be obtained from dimensional  reduction of six-dimensional ${\cal N}=(1,0)$
supergravity coupled to $\tilde{n}-1$ self-dual ${\cal N}=(1,0)$ tensor multiplets. Since the interacting non-abelian theory of ${\cal N}=(1,0)$ tensor multiplets is not known, and assuming that such a theory exists, our construction cannot be applied directly to calculating its amplitudes. For example, since this interacting tensor theory has no vector multiplets it cannot be constructed  in terms of scalar-coupled 
gauge theories. Rather, one may realize chiral tensor fields as bispinors and thus the relevant gauge theories should 
contain additional fermions \cite{Johansson:2014zca}.

The fact that two very different-looking theories in $D=6$ reduce to the same five-dimensional MESGT, belonging to the generic Jordan family, 
has a counterpart in theories with 16 supercharges. Namely, 
the $\cN=4$ sYM supermultiplet   
can be obtained from  ${\cal N}=(1,1)$ sYM multiplet as well as ${\cal N}=(2,0)$ tensor multiplet in six dimensions.
MESGTs describing the coupling of $n$  ${\cal N}=(1,1)$ vector multiplets to ${\cal N}=(1,1)$ 
supergravity  in six dimensions have the $U$-duality  groups  $SO(n,4)\times SO(1,1)$ and we  
expect the method presented in this paper to extend in a straightforward manner to the construction of the amplitudes of the 
corresponding ${\cal N}=(1,1)$  YMESGTs.  The formulation of a consistent interacting 
non-abelian theory of ${\cal N}=(2,0)$ multiplets coupled to ${\cal N}=(2,0)$ supergravity remains a fascinating open 
problem.\footnote{Note that the double-copy construction of the amplitudes of six-dimensional ${\cal N}=(2,2)$ Poincar\'e supergravity  in terms of ${\cal N}=(1,1)$ sYM amplitudes is well understood. 
On the level of spectra, the double-copy construction applied to the ${\cal N}=(2,0)$ theory is expected to give the amplitudes of ${\cal N}=(4,0)$ 
supergravity~\cite{Chiodaroli:2011pp} in six dimensions. The formulation of an interacting ${\cal N}=(4,0)$ 
supergravity in six dimensions is an open problem as well.}

\bigskip

\section*{Acknowledgements} 
We would like to thank Z.~Bern, J.J.~Carrasco, A.~Ochirov and A.~Tseytlin for useful discussions as well as collaborations on related topics.
One of us (M.G.) would like to thank the CERN Theory Division and the 
Albert Einstein Institute for their hospitality where part of this work was done. The  research of M.G. is 
supported in part by the US Department of Energy  under  DOE Grant No: DE-SC0010534. The work 
of R.R. is supported by the US Department of Energy under contract DE-SC0008745.

\newpage
\appendix

\renewcommand{\theequation}{A.\arabic{equation}}
\setcounter{equation}{0}

\section{Notation \label{appendixA}}
In this appendix we present a summary of the various indices used throughout the paper.    
\begin{center}
\begin{tabular}{ll}
 $\tilde n $ &  number of matter vector multiplets in $5D$.  \\[3pt]
 $K $ & compact subgroup of the isometry group $G$ \\
 & promoted to gauge symmetry. \\[3pt]   
$ A,B,C=-1,0, \ldots, \tilde  n$  & index running over all vector fields in $4D$. \\[3pt]
$ I,J,K=0, \ldots, \tilde  n$  & index running over matter vectors/multiplets in $4D$; \\
&index running over all vector fields in $5D$. \\[3pt]
$ a,b,c=1,2,\ldots, \tilde n$ & index running over extra vector fields in $4D$;\\
& flat target space index in $5D$.  \\[3pt]
$ x,y = 1,2, \ldots  , \tilde n $ & curved target space indices in $5D$.   \\[3pt]
$ m,n =1,2, \ldots , \tilde n-\text{dim}(K)$ &  index running over the extra  spectator fields in $4D$.   \\[3pt]
$ r,s,t=\tilde n+1 - \text{dim}(K), \ldots , \tilde n$ & index running over the non-abelian gauge fields.   \\[3pt]
$ \hat a , \hat b, \hat c $   &  gauge adjoint indices.   \\[3pt]
$ \hat \imath, \hat \jmath = 1,2$ & $SU(2)_R$ indices in $5D$ and $4D$.   \\[3pt]
\end{tabular}\\
\end{center}
With this notation we have 
\be A =\big( -1, \ I \big) = \big(-1, 0, \ a \big) = \big( -1, 0, \ m, \ r \big) \ .  \ee
In the YMESGT Lagrangians
we use the quantities $f^{rst}$ and $g$ to denote structure constants and coupling constant for the supergravity gauge group. 
These should not be confused with $\mathrm{f}^{\hat a \hat b \hat c}$ and $\mathrm{g}$, which denote structure constants and coupling constant
for the two gauge-theory factors employed in the double-copy construction. 
Additionally, $F^{rst}$ are the structure constants of the \emph{global} symmetry group of the cubic scalar couplings that are introduced in the 
$\cN=0$ gauge-theory factor. $g'$ is a proportionality constant that appears in the above couplings.

\renewcommand{\theequation}{B.\arabic{equation}}
\setcounter{equation}{0}

\section{Expansions for the generic Jordan family\label{appB}}

In this appendix we collect expansions for the period matrix, scalar metric and K\"{a}hler potential in the 
symplectic frame specified in section \ref{sec4dsymp}. All quantities are expanded up to terms quadratic in the scalar fields.
The period matrix $\cN_{AB}$ has the following entries,
\bea
{\cN}_{-1-1} &=& -  i + 2 i (z^I)^2 \ , \\
{\cN}_{-10} &=&  2  z^0 + 2 i (z^0)^2 + 2 i |z^a|^2 \ , \\
{\cN}_{-1a} &=&  2 (1+  \sqrt{2} z^1) z^a + 2 i z^0 \bar z^a + 2 i \bar z^0 z^a - \delta_{1a} \sqrt{2} (z^b)^2 \ ,  \eea
\bea
{\cN}_{00} &=& -  i + 2 i (z^0)^2 +  2 i (\bar z^a)^2 \ , \\
{\cN}_{0a} &=&  2 (1+  \sqrt{2} \bar z^1) \bar z^a + 2 i z^0  z^a + 2 i \bar z^0 \bar z^a - \delta_{1a} \sqrt{2} (\bar z^b)^2 \ ,  \\
{\cN}_{ab} &=& -(i - 2 \bar z^0) \delta_{ab} + 2 i z^a z^b + 2 i \bar z^a \bar z^b  \ .
\eea
The K\"{a}hler potential is given by
\be e^{{\cal K}} = {1\over 2} + {i \over 2} (z^0- \bar z^0) + {1 \over \sqrt{2} } (z^1 + \bar z^1) - {1 \over 2} (z^0 - \bar z^0)^2  + 
{i\over \sqrt{2} } (z^0 - \bar z^0) (z^1 + \bar z^1) + {3 \over 4} (z^1 + \bar z^1)^2 - {1 \over 4} (z^r - \bar z^r)^2 \ .  \ee
The scalar manifold metric has the following non-zero entries,
\bea
g_{00} &=& 1 +  2 i (z^0 - \bar z^0) - 3 (z^0 - \bar z^0)^2 \ , \\
g_{11} &=& 1 +  \sqrt{2} (z^1 + \bar z^1) + {3 \over 2} (z^1 + \bar z^1)^2 - {3 \over 2 }(z^t - \bar z^t)^2 \ , \\
g_{1r} &=& i \big( \sqrt{2}  + 3 z^1 + 3 \bar z^1 \big) (z^r - \bar z^r ) \ ,\\
g_{rs} &=& \Big( 1 +   \sqrt{2} (z^1 + \bar z^1) + {3 \over 2} (z^1 + \bar z^1)^2 - {1 \over 2 }(z^t - \bar z^t)^2 \Big) \delta_{rs} - (z^r - \bar z^r)(z^s - \bar z^s)
\ . \quad
\eea
These expansions are employed together with the Lagrangian (\ref{L4dfinal}) to derive the Feynman rules of the theory 
and to compare the amplitudes from the  supergravity Lagrangian with the ones obtained from the double-copy construction.

\end{document}